\documentclass[10pt,journal,comsoc]{IEEEtran}
\usepackage{cite}
\usepackage{bm}
\usepackage{amsmath}
\usepackage{extarrows}
\usepackage{amssymb}
\usepackage{graphicx}
\usepackage{color}
\usepackage{enumerate}
\usepackage{indentfirst}
\usepackage{bookmark} 
\graphicspath{{figure}}
\usepackage{setspace}
\usepackage{subfigure}
\usepackage{algorithm}
\usepackage{algorithmicx}
\usepackage{multirow}
\usepackage{stfloats}
\usepackage{nicematrix}
\usepackage{pifont}
\usepackage{threeparttable}
\usepackage{verbatim}

\newcommand{\mr}{\mathrm} 

\newcommand{\BE}{\begin{equation}}
\newcommand{\EE}{\end{equation}}
\newcommand{\BS}{\begin{subequations}}
\newcommand{\ES}{\end{subequations}}
\renewcommand{\bf}{\bm}

\newtheorem{theorem}{Theorem}

\newtheorem{definition}{Definition}

\newtheorem{lemma}{Lemma}

\allowdisplaybreaks \allowdisplaybreaks[2]

\ifCLASSINFOpdf
\graphicspath{{figure/}}

\else

\fi

\begin{document}
\title{{Memory AMP for Generalized MIMO: Coding Principle and Information-Theoretic Optimality}}

\author{{\IEEEauthorblockN{Yufei Chen, \emph{Student Member, IEEE}, Lei Liu,  \emph{Senior Member, IEEE},  Yuhao Chi, \emph{Member, IEEE},\\ Ying Li,  \emph{Member, IEEE}, and Zhaoyang Zhang, \emph{Senior Member, IEEE}}}

\thanks{Yufei Chen, Yuhao Chi, and Ying Li are with the State Key Laboratory of Integrated Services Networks, Xidian University, Xi'an, 710071, China (e-mail: yfchen1@stu.xidian.edu.cn, yhchi@xidian.edu.cn, yli@mail.xidian.edu.cn).}

\thanks{Lei Liu and Zhaoyang Zhang are with the Zhejiang Provincial Key Laboratory of Information Processing, Communication and Networking, College of Information Science and Electronic Engineering, Zhejiang University, Hangzhou 310007, China. (e-mail: lei\_liu@zju.edu.cn, ning\_ming@zju.edu.cn).}

}

\maketitle

\begin{abstract}
    To support complex communication scenarios in next-generation wireless communications, this paper focuses on a generalized MIMO (GMIMO) with practical assumptions, such as massive antennas, practical channel coding, arbitrary input distributions, and general right-unitarily-invariant channel matrices (covering Rayleigh fading, certain ill-conditioned and correlated channel matrices). The orthogonal/vector approximate message passing (OAMP/VAMP) receiver has been proved to be information-theoretically optimal in GMIMO, but it is limited to high-complexity linear minimum mean-square error (LMMSE). To solve this problem, a low-complexity memory approximate message passing (MAMP) receiver has recently been shown to be Bayes optimal but limited to uncoded systems. Therefore, how to design a low-complexity and information-theoretically optimal receiver for  GMIMO is still an open issue. To address this issue, this paper proposes an information-theoretically optimal MAMP receiver and investigates its achievable rate analysis and optimal coding principle. Specifically, due to the long-memory linear detection, state evolution (SE) for MAMP is intricately multi-dimensional and cannot be used directly to analyze its achievable rate. To avoid this difficulty, a simplified single-input single-output (SISO) variational SE (VSE) for MAMP is developed by leveraging the SE fixed-point consistent property of MAMP and OAMP/VAMP. The achievable rate of MAMP is calculated using the VSE, and the optimal coding principle is established to maximize the achievable rate. On this basis, the information-theoretic optimality of MAMP is proved rigorously. Furthermore, the simplified SE analysis by fixed-point consistency is generalized to any two iterative detection algorithms with the identical SE fixed point.
    Numerical results show that the finite-length performances of MAMP with practical optimized low-density parity-check (LDPC) codes are $0.5 \sim 2.7$~dB away from the associated constrained capacities. It is worth noting that MAMP can achieve the same performances as OAMP/VAMP with $4\text{\textperthousand}$ of the time consumption for large-scale systems.

\end{abstract}


\section{Introduction}
With the rapid development of wireless communications, 6G networks are expected to provide performance superior to 5G and satisfy emerging services and applications\cite{6G1,spectrumchain2023}, such as extended reality services \cite{VR3}, multimedia communications~\cite{YuhaoTMM}, and mobile unmanned aerial vehicle systems~\cite{UAV}. Accordingly, data types in various application scenarios become more diverse, and practical communication scenarios are more complex. However, most conventional multiple-input multiple-output (MIMO) technologies are limited to ideal communication assumptions, i.e., a limited number of antennas, no coding constraint, Gaussian signaling, channel state information (CSI) available at the transceiver, and independent identically distributed (IID) channel matrices, which cannot effectively support the complex 6G scenarios. Therefore, a more practical generalized MIMO (GMIMO)\cite{YuhaoTcom2022} is considered in this paper, including: 1) massive antennas, 2) practical channel coding and decoding, 3) arbitrary input distributions, 4) CSI only available at the receiver, and 5) general right-unitarily-invariant channel matrices, covering Rayleigh fading, certain ill-conditioned and correlated channel matrices. Meanwhile, these generalized assumptions bring new challenges to the design of receivers for GMIMO.

\subsection{Conventional Turbo Receivers}
The iterative linear minimum mean-square error receiver, termed Turbo-LMMSE, has been proved to achieve Gaussian capacity region of multi-user (MU) MIMO with Gaussian signaling~\cite{LeiTSP2019,YuhaoTWC2018}. Limited by the high-complexity LMMSE, Turbo-LMMSE is difficult to apply effectively to large-scale systems. To address this issue, a low-complexity Gaussian message passing algorithm is proposed and guaranteed to converge to LMMSE performance\cite{LeiTWC2016,LeiTWC2019}. However, these methods\cite{LeiTSP2019,YuhaoTWC2018,LeiTWC2016,LeiTWC2019} are limited to ideal Gaussian signaling and strictly suboptimal for practical discrete signaling, such as quadrature phase-shift keying (QPSK) and quadrature amplitude modulation (QAM).

\subsection{Advanced AMP-Type Receivers}
To solve the difficulty in conventional Turbo receivers, approximate message passing (AMP)-type receivers have been rapidly developed and widely used in MIMO receivers~\cite{MaTWC2019,Jintwc2020,ampdec1,MAMPOTFSconf}. AMP is a high-efficient signal recovery algorithm with a low-complexity matched filter (MF) for arbitrary input distributions \cite{donoho2009message,BayatiTIT2011}. Remarkably, the asymptotic performance of AMP can be evaluated using a scalar recursion called state evolution (SE)\cite{BayatiTIT2011}, based on which AMP is proved to be Bayes optimal\cite{ReevesTIT2019,Barbier2018b}. 
However, AMP is only available for IID channel matrices. For more complex non-IID channel matrices, AMP performs poorly or even diverges~\cite{Vila2015ICASSP,manoel2014sparse,Rangan2017TIT}. This severely restricts the application of AMP in more practical systems. 
To address the limitation of AMP, orthogonal/vector AMP (OAMP/VAMP) is developed in~\cite{MaAcess2017,Rangan2019TIT} for a wide range of right-unitarily-invariant matrices, employing an LMMSE to mitigate linear interference and orthogonalization to overcome the correlation problem during iteration. The Bayes optimality of OAMP/VAMP is proved via the replica methods in~\cite{Kabashima2006, Tulino2013TIT}, and lately rigorously for a general class of signal priors and arbitrary unitarily-invariant measurement matrix $\bf{A}$ under a ``high-temperature” condition that restricts the range of eigenvalues of $\bf{A}^T\bf{A}$ in~\cite{li2022random,Barbier2018ISIT}.

Although the above AMP-type algorithms \cite{donoho2009message,MaAcess2017,Rangan2019TIT} are Bayes optimal, they are limited to uncoded systems and ignore the effects of channel coding and decoding, leaving no guarantee of error-free signal recovery. In coded systems, the achievable rate is a key measurement with asymptotically error-free recovery.
Thus, the maximum achievable rate is commonly used to denote the information-theoretic (i.e., constrained-capacity) optimality of a coded system with fixed input distribution. For GMIMO with IID channel matrices and arbitrary input signaling, the achievable rate analysis and constrained-capacity optimality of AMP are presented in~\cite{LeiTIT2021} based on the scalar SE. Specifically, the optimal coding principle is derived via the perfect matching criterion between the single-input single-output (SISO) SE transfer functions of linear detector (LD) and nonlinear detector (NLD). Based on the mutual information-MMSE (I-MMSE) lemma\cite{GuoTIT2005}, the maximum achievable rate of AMP is proved to be equal to the associated constrained capacity \cite{LeiTIT2021}. For general right-unitarily-invariant channel matrices, OAMP/VAMP is shown to achieve the constrained capacity of point-to-point (P2P) GMIMO in~\cite{LeiOptOAMP} and the constrained-capacity region of MU-GMIMO in~\cite{YuhaoTcom2022}, respectively. Unlike AMP, since the orthogonalizations in LD and NLD destroy the minimum mean-square error (MMSE)  property, it is difficult to calculate the achievable rate of OAMP/VAMP directly based on I-MMSE lemma. To overcome this difficulty, a variational SE (VSE) of OAMP/VAMP is developed by incorporating all orthogonal operations into the LD\cite{LeiOptOAMP}, based on which the achievable rate analysis and optimal coding principle can be derived using I-MMSE lemma. However, due to the high complexity of the LMMSE, OAMP/VAMP cannot be used effectively for large-scale systems.

\subsection{Challenges}
The information-theoretic optimality of Turbo-LMMSE, AMP, and OAMP/VAMP receivers is restricted to Gaussian signaling, IID channel matrices, or high complexity, respectively, which are difficult to apply effectively to large-scale GMIMO. How to design a low-complexity and information-theoretic optimal receiver for GMIMO is still an open issue.

A promising low-complexity candidate receiver is memory AMP (MAMP), which is proved to be Bayes optimal for uncoded linear systems with arbitrary input signaling and right-unitarily-invariant matrices \cite{MAMPTIT}. Since MAMP employs a long-memory MF (LM-MF) to replace the LMMSE of OAMP/VAMP, its complexity is substantially lower than that of OAMP/VAMP and comparable to that of AMP. Because of the correlation between the long-memory inputs, the covariances between all input estimations need to be calculated in the SE of MAMP to evaluate the asymptotic performance. As a result, the SE transfer functions of MAMP are multi-dimensional, as opposed to the SISO SE transfer functions of AMP and OAMP/VAMP, which simply track the update of estimated variances\cite{BayatiTIT2011,MaAcess2017,Rangan2019TIT}.

Due to the intricately multi-dimensional SE of MAMP, the existing achievable rate analysis and optimal coding principles derived via SISO transfer functions\cite{LeiTIT2021,LeiOptOAMP,YuhaoTcom2022} are infeasible to directly extend to MAMP. In addition, the Bayes optimal MAMP directly combined with a well-designed P2P channel code still brings a large bit-error rate (BER) performance loss, which has been demonstrated in numerical results in this paper. As a result, designing a low-complexity and information-theoretically optimal MAMP receiver is challenging and of great practical importance.

\subsection{Contributions}
To address the above issue, we propose an information-theoretically optimal MAMP receiver and study its achievable rate and optimal coding principle. To avoid the complex multi-dimensional SE analysis of MAMP, a simplified SISO VSE for MAMP is derived by leveraging the SE fixed-point consistent lemma of MAMP and OAMP/VAMP. According to this SISO VSE, the achievable rate of MAMP is obtained using I-MMSE lemma. Subsequently, the optimal coding principle of MAMP is derived with the goal of maximizing the achievable rate, and the maximum achievable rate equals the constrained capacity of GMIMO. Therefore, the constrained-capacity optimality of MAMP is established. Meanwhile, a more general theoretical-analysis equivalence theorem is provided, indicating that two arbitrary iterative detection algorithms with the identical SE fixed point have the same achievable rates and optimal coding principle. 
Moreover, taking the ill-conditioned and correlated channel matrices as examples, we compare the maximum achievable rates of MAMP and the existing cascading MAMP (CAS-MAMP, i.e., with separate MLD and NLD). Furthermore, a kind of capacity-approaching low-density parity-check (LDPC) code is designed for MAMP in GMIMO with Rayleigh fading, ill-conditioned, and correlated channel matrices, respectively. The main contributions of this paper are summarized as follows.
\begin{enumerate}
    \item  A simplified SISO VSE for MAMP is proposed to analyze its achievable rate, and the optimal coding principle is derived with the aim of maximizing the achievable rate.
    \item The constrained-capacity optimality of MAMP is proved, i.e., the maximum achievable rate of MAMP is equal to the constrained capacity of GMIMO.
    \item A general equivalence theorem is provided to validate that any two iterative detection algorithms with the identical SE fixed point have the same achievable rates and optimal coding principle.
     \item A kind of practical LDPC code is designed for MAMP, whose theoretical decoding thresholds are about $0.3$ dB away from the associated constrained capacities. Numerical results show that the finite-length performances of MAMP with optimized LDPC codes are significantly superior to those of MAMP with well-designed P2P LDPC codes. Furthermore, MAMP only takes $4\text{\textperthousand}$ of the execution time of the state-of-the-art OAMP/VAMP to achieve the same performances in large-scale systems.
\end{enumerate}

In summary, this is the first work to provide a low-complexity and constrained-capacity optimal receiver for coded GMIMO, which is also the first time to apply MAMP to GMIMO. For simplicity, this paper considers P2P-GMIMO, termed GMIMO, which can be directly extended to MU-GMIMO similarly as \cite{YuhaoTcom2022}.

\subsection{Connection to Existing Works}
\subsubsection{Other related low-complexity AMP-type receivers}
Recently, a low-complexity convolutional AMP (CAMP) has been proposed in~\cite{Takeuchi2020CAMP}, which replaces the Onsager term of AMP with a convolution of all preceding messages. In contrast to MAMP, CAMP converges slowly and even easily diverges for the channel matrices with high-condition numbers, making it difficult to apply to complex communication scenarios. In addition, a sufficient statistic MAMP (SS-MAMP) has been proposed in \cite{SS-MAMP} to accelerate the convergence of MAMP in uncoded systems.
Therefore, the theoretical results of this paper can be extended to SS-MAMP as a candidate for a constrained-capacity optimal receiver with faster convergence.
\subsubsection{Differences from the Turbo-LMMSE receiver in \texorpdfstring{\cite{LeiTSP2019,YuhaoTWC2018}}{}}
The Turbo-LMMSE receiver has been proved to be capacity-optimal for MIMO systems with ideal Gaussian signaling. In contrast, the constrained-capacity optimal MAMP receiver proposed in this paper are available for GMIMO systems with arbitrary signaling distribution.
\subsubsection{Differences from the capacity optimality of AMP and OAMP/VAMP receivers in \texorpdfstring{\cite{LeiTIT2021,LeiOptOAMP,YuhaoTcom2022}}{}}
The constrained-capacity optimality of AMP with low complexity is limited to IID channel matrices. 
Although OAMP/VAMP has been proved to be constrained-capacity optimal for more general right-unitarily-invariant channel matrices, it is difficult to apply to large-scale systems due to high-complexity LMMSE.
In contrast to AMP and OAMP/VAMP, the proposed MAMP is low-complexity and proved to be constrained-capacity optimal for right-unitarily-invariant channel matrices in this paper.
Furthermore, different from the simple SISO SE of AMP and OAMP/VAMP, the SE of MAMP is intricately multi-dimensional, making it infeasible to directly extend the existing analytical methods to MAMP.

\subsubsection{Differences from the Bayes optimal MAMP receiver in~\texorpdfstring{\cite{MAMPTIT}}{}}
The Bayes optimal MAMP receiver only focuses on uncoded systems, ignoring the effect of channel coding and decoding with
no guarantee of asymptotically error-free recovery. 
Furthermore, since the SE of MAMP is multi-dimensional, it cannot be utilized directly to analyze the achievable rate and optimal coding principle of MAMP.
To overcome this difficulty, a simplified SISO VSE for MAMP is proposed in this paper to derive its achievable rate and optimal coding principle. On this basis, the information-theoretic optimality of MAMP is proved rigorously.  
The numerical results of this paper demonstrate that the proposed MAMP with practical optimized LDPC codes can achieve capacity-approaching performance, while the Bayes optimal MAMP with well-designed P2P LDPC codes still brings a significant performance loss.

\subsection{Notations}
Bold uppercase (lowercase) letters denote matrices (column vectors).
$[\cdot]^{\rm{T}}$, $[\cdot]^{\rm{H}}$, and $[\cdot]^{-1}$ denote the transpose, conjugate transpose, and inverse operations, respectively. 
$\bf{I}$ and $\bf{0}$ are identity matrix and zero matrix or vector. 
$\rm min(\mathcal{S})$, $\rm max(\mathcal{S})$, and $|\mathcal{S}|$ denote the the minimum value, maximum value, and cardinality of set $\mathcal{S}$, accordingly. 
Denote $\|\bf{a}\|$ for the $\ell_2$-norm of vector $\bf{a}$, $\mr{tr}(\bf{A})$ for the trace of matrix $\bf{A}$, $(\cdot)^{\tau}$ for the $\tau$-th power of the value or matrix in the parentheses, $\mr{E}\{\cdot\}$ for the expectation over all random variables included in the brackets, $\mr{E}\{a|b\}$ for the expectation of $a$ conditional on $b$, $\rm{mmse}\{a|b\}$ for $\mr{E}\{(a-E\{a|b\})^2|b\}$, $\mathcal{CN}(\bf{\mu},\bf{\Sigma})$ for the circularly-symmetric Gaussian distributions with mean $\bf{\mu}$ and covariance $\bf{\Sigma}$, and $\left \langle \bf{A}_{M \times N} \mid \bf{B}_{M \times N} \right \rangle \equiv \tfrac{1}{N}\bf{A}^{\rm{H}}_{M \times N}\bf{B}_{M \times N}$. $X \sim Y$ represents that $X$ follows the distribution $Y$.
$\overset{\text{a.s.}}{=}$ denotes almost sure equivalence. 
A matrix is said to be column-wise independent identically distributed Gaussian (IIDG) and row-wise joint-Gaussian (CIIDG-RJG) if its each column is IIDG and its each row is joint Gaussian.

\subsection{Paper Outline}
This paper is organized as follows. Section II presents the system model of GMIMO. Section~III proposes an MAMP receiver in coded GMIMO. The coding principle and information-theoretic optimality of MAMP are derived in Section IV. Numerical simulations are provided in Section~V and the conclusion is presented in Section VI.

\section{System Model and Challenges}
In this section, the model and assumptions of GMIMO systems are provided. Then, the key challenges of GMIMO receiver design are presented.

\begin{figure*}[!tbp]
	\centering
	\includegraphics[width=0.9\linewidth]{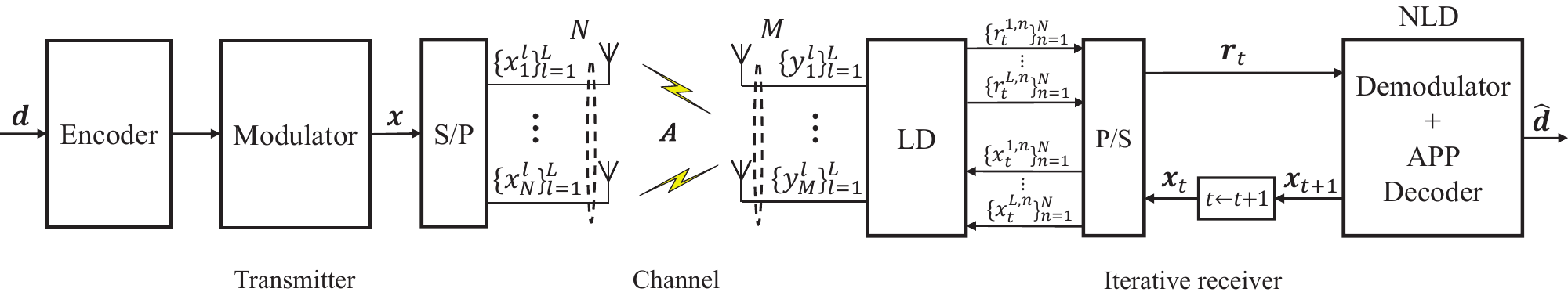}
	\caption{An uplink GMIMO system: an $N$-antennas transmitter and an $M$-antennas iterative receiver consisting of an LD and an NLD. S/P and P/S denote serial-to-parallel and parallel-to-serial conversion, respectively.}\vspace{0.2cm}\label{Fig:System model} 
\end{figure*}

\subsection{System Model}
Fig.~\ref{Fig:System model} illustrates an uplink GMIMO system with an $N$-antennas transmitter and one receiver equipped with $M$ antennas. In the transmitter, message sequence $\bf{d}$ is encoded by a forward error control (FEC) encoder. After modulation, length-$NL$ modulated sequence $\bf{x}$ is generated and transformed into $N$ sequences $\{x^l_{n}\}_{l=1}^L, n=1,...,N,$ by serial-to-parallel conversion, in which each entry of $\bf{x}$ is taken from a discrete constellation set $\cal{S}$. At the $l$-th time slot, symbol sequence $\bf{x}^l=[x^l_{1},...,x^l_{N}]^{\rm{T}}$ is transmitted to the channel,  satisfying the power constraint $\tfrac{1}{N}\mr{E}\{\|\bf{x}^l\|^2\}=1$.

The receiver obtains signal $\bf{y}^l= [y_1^l,...,y_M^l]^{\rm{T}}$ given by
\BE\label{Eqn:y^t}
\bf{y}^l=\bf{A}\bf{x}^l+\bf{n}^l,\;\;  l=1,\dots,L,
\EE
where $\bf{A}\in \mathbb{C}^{M\times N}$ is a channel matrix and $\bf{n}^{l} \sim\mathcal{CN}(\mathbf{0},\sigma^2\bm{I})$ is an additive white Gaussian noise (AWGN) vector. Without loss of generality, we assume $\frac{1}{\mathcal{F}}{\rm tr}\{\bf{A}^{\rm{H}}\bf{A}\}=1$ with $\mathcal{F}={\rm max} \{ M,N \} $, and define the signal-to-noise ratio (SNR) as ${snr} = \sigma^{-2}$.

Based on $\bf{y}^l$, an iterative receiver is implemented to recover message sequence $\bf{d}$, which consists of an LD and an NLD. The LD corresponds to the linear constraint~\eqref{Eqn:y^t}, and the NLD, consisting of a demodulator and an \emph{a-posteriori probability} (APP) decoder, corresponds to the FEC coding constraint $\bf{x} \in \mathcal{C}$ ($\mathcal{C}$ is the set of codewords). To be specific, as shown in Fig.~\ref{Fig:System model}, based on $\bf{y}^l$ and $\bf{x}^l_t=[x^{l,1}_t,\cdots,x^{l,N}_t]^{\rm{T}}$, the output estimation of LD is $\bf{r}^l_t=[r^{l,1}_t,\cdots,r^{l,N}_t]^{\rm{T}}$. After parallel-to-serial conversion, $\bf{r}_t=[\bf{r}^{1^{\rm{T}}}_t,\cdots,\bf{r}^{L^{\rm{T}}}_t]^{\rm{T}}$ is input to the NLD at $t$-th iteration and then the updated estimation $\bf{x}_{t+1}=[\bf{x}^{1^{\rm{T}}}_{t+1},\cdots,\bf{x}^{L^{\rm{T}}}_{t+1}]^{\rm{T}}$ is fed back to the LD.
The iterative process stops when message sequence $\hat{\bf{d}}$ is recovered successfully or the maximum number of iterations is reached.

\subsection{Assumptions and Challenges of GMIMO}
\subsubsection{Assumptions}
The GMIMO system satisfies the following assumptions.
\begin{itemize}
	\item There are a large number of transmit antennas and receive antennas, i.e., $N,M \rightarrow \infty$ and channel load $\beta=N/M$ is fixed.
	\item The entries of signal $\bf{x}$ are taken from an arbitrary distribution (e.g., QPSK, QAM, Gaussian, Bernoulli-Gaussian, etc.).
	\item Channel matrix $\bf{A}$ is right-unitarily-invariant, covering various types of channel matrices, e.g., IID random matrices (i.e., Rayleigh fading matrices), certain ill-conditioned and correlated matrices, which can be utilized to characterize complex practical communication scenarios\footnote{In practice, MAMP is not limited to right-unitarily-invariant channel matrices. For a given channel matrix, if the simulated MSE of MAMP can be accurately predicted by SE, the proposed achievable rate analysis and coding principle are also applicable to this given channel matrix.}~\cite{YuhaoTcom2022,MaTWC2019,Poor2021TWC}. Let the SVD of $\bf{A}$ be $\bf{A} = \bf{U}\bf{\Lambda}\bf{V}^{\rm{H}}$, where $\bf{U} \in \mathbb{C}^{M\times M}$ and $\bf{V} \in \mathbb{C}^{N\times N}$ are unitary  matrices, and $\bf{\Lambda}\in \mathbb{C}^{M\times N}$ is a rectangular diagonal matrix. $\bf{U}\bf{\Lambda}$ and $\bf{V}$ are independent, and $\bf{V}$ is Haar-distributed (uniformly distributed over all unitary matrices) \cite{RandomWire}. 
    \item Channel matrix $\bf{A}$ is only available to the receiver but unknown to the transmitter.
\end{itemize}	

\subsubsection{Challenges}
The above general assumptions bring some new challenges to the design of GMIMO receiver.

\begin{itemize}
\item How to achieve the information-theoretically optimal performance of GMIMO with low complexity is still an open issue. Note that the OAMP/VAMP receiver has been proved to be constrained-capacity optimal for P2P-GMIMO in\cite{LeiOptOAMP} and for MU-GMIMO in\cite{YuhaoTcom2022}, respectively. However, in the LD of OAMP/VAMP, a high-complexity LMMSE detection is utilized, making it difficult to apply effectively for practical large-scale systems. To address this issue, MAMP is proposed recently in~\cite{MAMPTIT}, which employs an alternative LM-MF in the LD to suppress linear interference. Meanwhile, MAMP has been shown to achieve Bayes optimality for uncoded systems with much lower complexity than OAMP/VAMP, but ignores the effect of channel coding and decoding.
\item The design principle of practical FEC codes for MAMP is still unclear. 
Since the memory is involved in local detectors, the SE of MAMP is intricately multi-dimensional. However, the existing analytical methods are all based on SISO SE (e.g., the achievable rate analysis and coding principle of AMP in \cite{LeiTIT2021} and OAMP/VAMP in \cite{LeiOptOAMP,YuhaoTcom2022}). Therefore, it is infeasible to extend the existing methods to analyze MAMP directly. As a result, how to obtain the coding principle of practical FEC codes for MAMP is still an open issue.
\end{itemize}

\section{MAMP Receiver and State Evolution}
In this section, we firstly introduce the construction of MAMP and then present the MAMP receiver and its state evolution for GMIMO.
\vspace{-0.3cm}
\subsection{MAMP Construction}
Since the detection process of \eqref{Eqn:y^t} in each time slot is the same, the time index $l$ is omitted in the rest of this paper for simplicity. Then, the received signal in \eqref{Eqn:y^t} can be rewritten as:
\BS\label{Eqn:y}
\begin{align}
\text{Linear constraint}\quad \Gamma:& \;\;\; \bf{y} = \bf{A}\bf{x} + \bf{n},\\
\text{Code constraint}\quad \Phi_{\mathcal{C}}: &\;\;\; \bf{x} \in \mathcal{C} \;\; {\mr{and}}\;\; x_i \sim P_X(x_i), \forall i.
\end{align}
\ES

Based on \eqref{Eqn:y}, an MAMP can be constructed based on a general memory iterative process (MIP) \cite{SS-MAMP} below.

\begin{definition}[Memory Iterative Process (MIP)]\label{Def:MIP}
An MIP involves a memory LD (MLD) and a memory NLD (MNLD) given by:
\BS\label{Eqn:MIP}
\begin{align}
    \mathrm{MLD}:  \;\;\;\;\;\; \bf{r}_t  &=  \gamma_{t}(\bf{X}_t)=\bf{\mathcal{Q}}_t\bf{y}+{\textstyle \sum_{i=1}^{t}}\bf{\mathcal{P}}_{t,i}\bf{x}_i,  \label{Eqn:MIP-MLD}\\
    \mathrm{MNLD}: \;\;\bf{x}_{t+1} &= \phi_t(\bf{R}_t),\label{Eqn:MIP-MNLD}
\end{align}
\ES
where $t$ starts from 1, $\bf{X}_t=[\bf{x}_1,...,\bf{x}_t]$, $\bf{R}_t=[\bf{r}_1,...,\bf{r}_t]$, and $\mathcal{\bf{Q}}_t\bf{A}$ and $\mathcal{\bf{P}}_{t,i}$ are polynomials in $\bf{A}^{\rm{H}}\bf{A}$.
Without loss of generality, the norms of $\mathcal{\bf{Q}}_t$ and $\mathcal{\bf{P}}_{t,i}$ are assumed as finite, such that $\gamma_{t}(\cdot)$ is Lipschitz-continuous~\cite{MAMPTIT}. Moreover, the conventional non-memory iterative process (e.g., AMP, OAMP/VAMP) is a special case of MIP, which includes $\gamma_{t}(\bf{x}_t)$ and $\phi_t(\bf{r}_t)$.
\end{definition}

Let $\bf{X}=\bf{x}\cdot \bf{1}^\mathrm{T}$ with an all-ones vector $\bf{1}$ of proper size and define the estimation errors $\bf{G}_{t}=[\bf{g}_1, ..., \bf{g}_t]$ and $\bf{F}_{t}=[\bf{f}_1, ..., \bf{f}_t]$ as
\BE\label{Eqn:MIP-error}
    \bf{G}_{t} = \bf{R}_{t} - \bf{X}, \;\;\;\;\;\;
    \bf{F}_{t} = \bf{X}_{t}- \bf{X},
\EE
with zero means and covariance matrices: 
\BE\label{Eqn:cor-matrix}
    \bf{V}^{\gamma}_{t} \equiv  \left \langle \bf{G}_{t} \mid \bf{G}_{t} \right \rangle , \;\;\;\;\;\;
    \bf{V}^{\phi}_{t} \equiv   \left \langle \bf{F}_{t} \mid \bf{F}_{t} \right \rangle .
\EE

\begin{definition}[MAMP]\label{def:MAMP}
The MIP is referred to as an MAMP when the following orthogonal constraints are satisfied: for $t \ge 1$, 
\BE\label{Eqn:Orth}
	 \left \langle \bf{g}_{t} \mid \bf{x} \right \rangle  \overset{\text{a.s.}}{=} 0, \;\;\;\;
	 \left \langle \bf{g}_{t} \mid \bf{F}_t \right \rangle  \overset{\text{a.s.}}{=} \bf{0},   \;\;\;\;
	 \left \langle \bf{f}_{t+1} \mid \bf{G}_t \right \rangle  \overset{\text{a.s.}}{=} \bf{0}.
\EE
\end{definition}

Different from a non-memory iterative process (e.g., OAMP/VAMP), the ``full orthogonality'' in \eqref{Eqn:Orth} is necessary for MAMP to make the current output estimation error orthogonal to all preceding input estimation errors in each estimation. Based on the orthogonality, the following lemma is given to ensure the accuracy of SE for MAMP.
\begin{lemma}[Asymptotically IID Gaussianity]\label{lem:IIDG} Assume that $\{\gamma_t(\cdot)\}$ is Lipschitz-continuous and $\{\phi_t(\cdot)\}$ is separable-and-Lipschitz-continuous. With the orthogonality in \eqref{Eqn:Orth}, the asymptotically IID Gaussianity of MAMP is given as follows: for $1\le t^{\prime}\le t$,
    \BS\label{Eqn:IIDG}\begin{align}
	v^{\gamma}_{t,t^{\prime}}  &\overset{\text{a.s.}}{=}     \langle \gamma_t(\bf{X}+\bf{Z}^{\phi}_t) - \bf{x} \mid \gamma_{t^{\prime}}(\bf{X}+\bf{Z}^{\phi}_{t^{\prime}}) - \bf{x}  \rangle     ,\\
	v^{\phi}_{t+1,t^{\prime}+1}  &\overset{\text{a.s.}}{=}    \langle \phi_t(\bf{X}+\bf{Z}^{\gamma}_t) - \bf{x} \mid \phi_{t^{\prime}}(\bf{X}+\bf{Z}^{\gamma}_{t^{\prime}}) - \bf{x}  \rangle     ,
	\end{align}
	\ES
	where $\{v^{\gamma}_{i,j}\}$ and $\{v^{\phi}_{i,j}\}$ are the elements of $\bf{V}^{\gamma}_{t}$ and $\bf{V}^{\phi}_{t}$ respectively, $\bf{Z}^{\gamma}_t=[\bf{z}^{\gamma}_1,...,\bf{z}^{\gamma}_t]$ and $\bf{Z}^{\phi}_t=[\bf{z}^{\phi}_1,...,\bf{z}^{\phi}_t]$ are CIIDG-RJG and independent of $\bf{x}$. Moreover,
	$\langle \bf{Z}^{\gamma}_t \mid \bf{Z}^{\gamma}_t  \rangle   = \bf{V}^{\gamma}_{t}$ and 
	$\langle \bf{Z}^{\phi}_t \mid \bf{Z}^{\phi}_t \rangle  = \bf{V}^{\phi}_{t}$. In detail, $\bf{z}^{\gamma}_t \sim\mathcal{CN}(\mathbf{0},v^{\gamma}_{t,t}\bm{I})$ with $\mr{E}\{\bf{z}^{\gamma}_t(\bf{z}^{\gamma}_{t^{\prime}})^{\rm{H}}\}=v^{\gamma}_{t,t^{\prime}}\bm{I}$ and $\bf{z}^{\phi}_t \sim\mathcal{CN}(\mathbf{0},v^{\phi}_{t,t}\bm{I})$ with $\mr{E}\{\bf{z}^{\phi}_t(\bf{z}^{\phi}_{t^{\prime}})^{\rm{H}}\}=v^{\phi}_{t,t^{\prime}}\bm{I}$.
\end{lemma}

Based on Definitions \ref{Def:MIP} and \ref{def:MAMP}, a construction for MAMP is presented in the following lemma.

\begin{lemma}[MAMP Construction]\label{lem:MAMP Construction}
Given a general $\hat{\gamma_t}(\cdot)$ below 
\BE
 \hat{\gamma}_{t}(\bf{X}_t)=\bf{Q}_t\bf{y}+{\textstyle \sum_{i=1}^{t}}\bf{P}_{t,i}\bf{x}_i, 
\EE
and an arbitrary differentiable, separable and Lipschitz-continuous $\hat{\phi}_t(\cdot)$, we can construct an MAMP as follows:
\BS\label{Eqn:MAMP equation}
\begin{align}
\mathrm{MLD}: \quad \gamma_{t}(\bf{X}_t)&= \tfrac{1}{\varepsilon^{\gamma}_t}(\hat{\gamma}_{t}(\bf{X}_t)-\bf{X}_t\bf{p}_t),\label{Eqn:gamma equation}\\
\mathrm{MNLD}: \quad \phi_{t}(\bf{R}_t)&= \tfrac{1}{\varepsilon^{\phi}_t}(\hat{\phi}_{t}(\bf{R}_t)-\bf{R}_t\bf{w}_t),\label{Eqn:phi equation}
\end{align}
\ES
where
\BS\label{Eqn:parameter}
\begin{align}
\varepsilon^{\gamma}_t &= \tfrac{1}{N}\mr{tr}\{\bf{Q}_t\bf{A}\},\label{Eqn:parameter1}\\
\bf{p}_t &= [\tfrac{1}{N}\mr{tr}\{\bf{P}_{t,1}\} \cdots \tfrac{1}{N}\mr{tr}\{\bf{P}_{t,t}\} ]^{\rm{T}},\label{Eqn:parameter2}\\
\bf{w}_t &= [\mr{E}\{\tfrac{\partial \hat{\phi}}{\partial r_1}\} \cdots \mr{E}\{\tfrac{\partial \hat{\phi}}{\partial r_t} \}]^{\rm{T}},\label{Eqn:parameter3}
\end{align}
\ES
and $\varepsilon^{\phi}_t$ is an arbitrary constant and generally determined by minimizing the MSE of $\phi_t(\cdot)$. Meanwhile, the normalized parameters $\{\varepsilon^{\gamma}_t\}$ and orthogonal parameters $\{\bf{p}_t, \bf{w}_t\}$ are designed to ensure the orthogonality in \eqref{Eqn:Orth}.
\end{lemma}

\begin{figure*}[!t]
    \centering
    \subfigure[MAMP receiver: $\hat{\gamma}_t(\cdot)$ and $\hat{\phi}_t(\cdot)$ denote the LM-MF detection and demodulation and APP decoding for the local constraints $\Gamma$ and $\Phi_{\mathcal{C}}$, respectively.]
    {
        \begin{minipage}[t]{0.72\linewidth}\label{Fig:MAMP receiver} 
            \centering
            \includegraphics[width=0.85\linewidth]{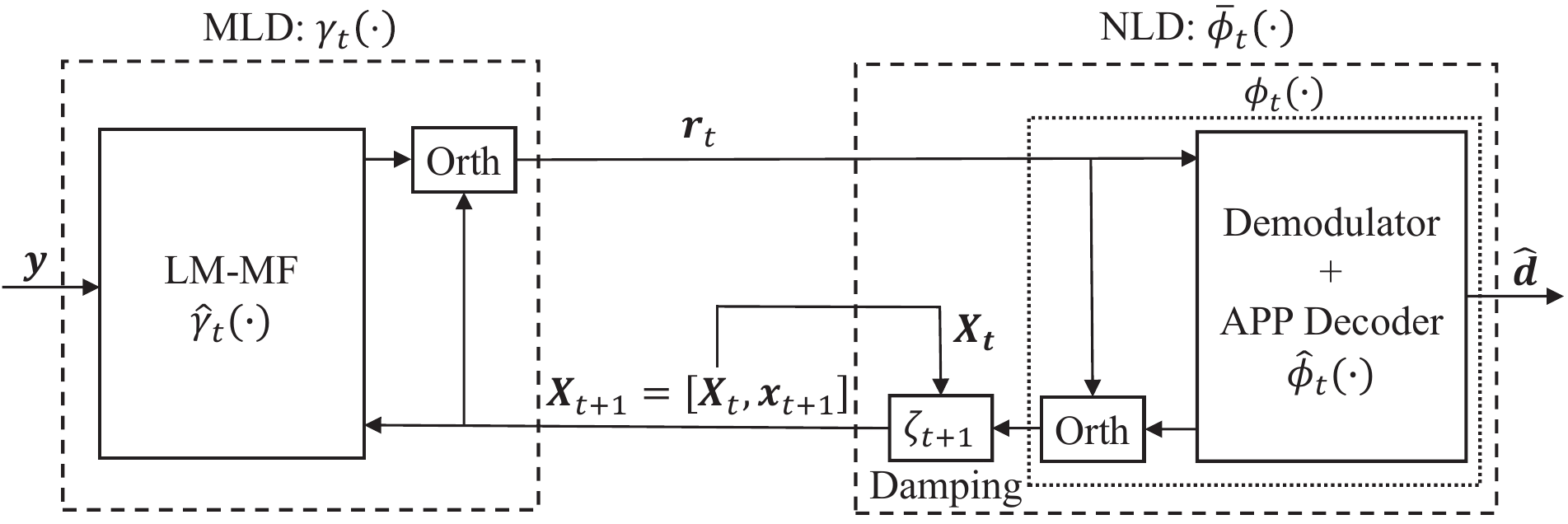}\vspace{0.3cm}
		\end{minipage}
    }\\
    
    \subfigure[Transfer functions: $\gamma_{\rm SE}$ and $\bar{\phi}_{\rm SE}$ are the MSE transfer functions of $\gamma_t$ and $\bar{\phi}_t$, respectively.]
	{
		\begin{minipage}[t]{0.72\linewidth}\label{Fig:MAMP SE} 
			\centering
			\includegraphics[width=0.7\linewidth]{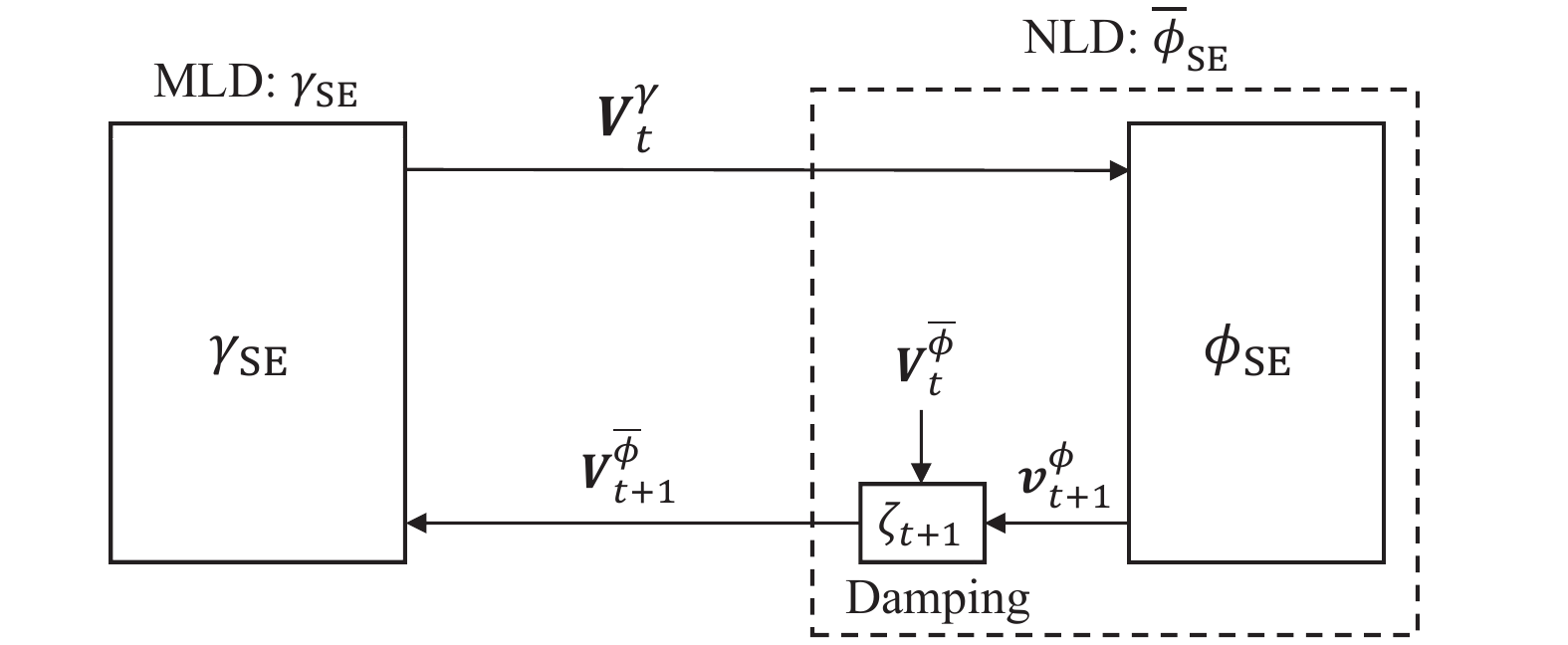}\vspace{0.3cm}
		\end{minipage} 
	}
	\caption{Illustration of the MAMP receiver and its state evolution.}\vspace{0.2cm}
\end{figure*}

\subsection{MAMP Receiver and State Evolution}
For coded GMIMO, $\hat{\phi}_t(\bf{R}_t)$ in \eqref{Eqn:phi equation} corresponds to the code constraint $\Phi_{\mathcal{C}}$. Note that the output of the FEC decoder depends only on current input estimations $\bf{r}_t$ at the $t$-th iteration. As a result, ${\phi}_t(\bf{R}_t)$ and local $\hat{\phi}_t(\bf{R}_t)$ in \eqref{Eqn:phi equation} are degraded to the following non-memory form:
\BE\label{Eqn:nonme_nld}
\phi_t(\bf{r}_t) \equiv \tfrac{1}{\varepsilon^{\phi}_t}(\hat{\phi}_t(\bf{r}_t)-w_t\bf{r}_t).
\EE
Meanwhile, a damping strategy is employed for the output of $\phi_t(\bf{r}_t)$ to guarantee and improve MAMP convergence while also preserving orthogonality, as demonstrated in \cite{MAMPTIT}. The detailed process of MAMP receiver in GMIMO is given as follows.

\subsubsection{MAMP Receiver}
Based on Lemma~\ref{lem:MAMP Construction} and \eqref{Eqn:nonme_nld}, Fig. \ref{Fig:MAMP receiver} shows the MAMP receiver for coded GMIMO, i.e., starting with $t=1$ and $\bf{X}_1=\bf{0}$,
\BS\label{Eqn:BO-MAMP}\begin{align}
	\!\!\mathrm{MLD:}\;\;\;\;\bf{r}_{t} &= \gamma_t(\bf{X}_t) = \tfrac{1}{\varepsilon^{\gamma}_t}(\hat{\gamma}_{t}(\bf{X}_t)-\bf{X}_t\bf{p}_t )  ,\label{Eqn:MAMP-MLE}\\
	\!\!\mathrm{NLD:}\;\bf{x}_{t+1} &= \bar{\phi}_t(\bf{r}_t) = [\bf{X}_t,\phi_t(\bf{r}_t)]\cdot \bf{\zeta}_{t+1},\label{Eqn:MAMP-NLE}
\end{align}  
\ES
where $\bf{X}_t=[\bf{x}_1,...,\bf{x}_t]$, and damping vector $\bf{\zeta}_{t+1} = [\zeta_{t+1,1},...,\zeta_{t+1,t+1}]^\mathrm{T}$ with $\textstyle \sum_{i=1}^{t+1}\zeta_{t+1,i}=1$ is utilized to optimize the linear combination of all the estimations $\{\bf{x}_t\}$.

In~\eqref{Eqn:MAMP-MLE}, the local LM-MF function of $\gamma_t(\bf{X}_t)$ is 
\BE\label{Eqn:localLMMF}
\hat{\gamma}_{t}(\bf{X}_t)=\bf{A}^{\rm{H}}\tilde{\gamma}_t(\bf{X}_t), 
\EE
with
\BE\label{Eqn:tilde-gamma}
\tilde{\gamma}_t(\bf{X}_t)=\theta_t \bf{B} \tilde{\gamma}_{t-1}(\bf{X}_{t-1}) + \xi_t(\bf{y}-\bf{A}\bf{x}_t),
\EE
where $\tilde{\gamma}_{0}(\bf{X}_{0})=\bf{0}$, $\bf{B}=\lambda^{\dagger}\bf{I}-\bf{A}\bf{A}^{\rm{H}}$ with $\lambda^{\dagger}=(\lambda_{\text{min}}+ \lambda_{\text{max}})/2$, $\lambda_{\text{min}}$ and $\lambda_{\text{max}}$ denote the minimal and maximal eigenvalues \footnote{The detailed discussion on finding $\lambda_{\text{min}}$ and $\lambda_{\text{max}}$ is presented in Subsection~\ref{sec:Bound}.} of $\bf{A}\bf{A}^{\rm{H}}$, and $\lambda^{\dagger}$ is utilized to calculate relaxation parameters $\{\theta_t\}$ to ensure and improve the convergence of the MAMP receiver\cite[Section V-A]{MAMPTIT}. In addition, weights $\{\xi_t\}$ can be optimized to further accelerate the convergence speed of MAMP \cite[Section V-B]{MAMPTIT}. 
Noting that when $t\to\infty$, $\tilde{\gamma}_t(\bf{X}_t) = \tilde{\gamma}_{t-1}(\bf{X}_{t-1}) = \xi_t(\bf{I}-\theta_t\bf{B})^{-1}(\bf{y}-\bf{A}\bf{x}_t)$ in \eqref{Eqn:tilde-gamma}, the output of $\hat{\gamma}_{t}(\bf{X}_t)$ converges to the LMMSE estimation in the LD of OAMP/VAMP\cite{LeiOptOAMP}, i.e.,
\BE\nonumber
\!\!\!\hat{\gamma}_{t}(\bf{X}_t) = \bf{A}^{\rm{H}}\tilde{\gamma}_t(\bf{X}_t) \to \tfrac{\xi_t}{\theta_t}\bf{A}^{\rm{H}}[(\tfrac{1}{\theta_t}-\lambda^{\dagger}) \bf{I}+\bf{A}\bf{A}^{\rm{H}}]^{-1}(\bf{y}-\bf{A}\bf{x}_t).
\EE

\vspace{-0.3cm}
In~\eqref{Eqn:MAMP-NLE}, the local MMSE-optimal estimation function of ${\phi}_t(\bf{r}_t)$ is given by
\BE\label{Eqn:decoder}
\hat{\phi}_t(\bf{r}_t)\equiv\mr{E}\{\bf{x} |\bf{r}_t,\Phi_{\mathcal{C}}\},
\EE
which corresponds to the demodulation and APP channel decoding \cite[Equation(10)]{MaTWC2019}. It is noted that $\hat{\phi}_t(\cdot)$ is assumed to be Lipschitz-continuous in this paper.

\subsubsection{State Evolution (SE)}\label{sec:SE}
Since the LM-MF is employed in MLD, a covariance-matrix SE is required to evaluate the asymptotic performance. 
Due to the right-unitarily-invariant property of $\bf{A}$, the IID Gaussianity property in Lemma~\ref{lem:IIDG} are valid, based on which the asymptotic MSE performance
of MAMP can be predicted by the MSE functions $\gamma_{\rm SE}(\cdot)$ and $\bar{\phi}_{\rm SE}(\cdot)$ in the SE, i.e.,
\BS\label{Eqn:SE}
\begin{align}
\mathrm{MLD:}\;\;\;\;\;\;\;\bf{V}^{\gamma}_{t} &= \gamma_{\rm SE}(\bf{V}^{\bar{\phi}}_{t}),\label{Eqn:SE-MLE}\\
\mathrm{NLD:}\;\;\;\;\;\bf{V}^{\bar{\phi}}_{t+1} &= \bar{\phi}_{\rm SE}(\bf{V}^{\gamma}_{t}),\label{Eqn:SE-NLE}
\end{align}
\ES
where $\bf{V}^{\gamma}_{t}$ and $\bf{V}^{\bar{\phi}}_{t}$ are the covariance matrices defined in \eqref{Eqn:cor-matrix}, and $\gamma_{\rm SE}(\cdot)$ and $\bar{\phi}_{\rm SE}(\cdot)$ correspond to constraints \eqref{Eqn:MAMP-MLE} and \eqref{Eqn:MAMP-NLE}, respectively. Moreover, Fig.~\ref{Fig:MAMP SE} gives a graphical illustration of the SE in \eqref{Eqn:SE}. 
Particularly, although the NLD is non-memory, the covariance matrix $\bf{V}_{t+1}^{\bar{\phi}}$ in \eqref{Eqn:SE-NLE} still needs to be calculated as the input of the MSE function of MLD, which is obtained by the damping operation of $\bf{V}_{t}^{\bar{\phi}}$ and $\bf{v}_{t+1}^{\phi}$, i.e., the $(t+1)$-th row of $\bf{V}_{t+1}^{\bar{\phi}}$ is
\BE
(\bf{v}^{\bar{\phi}}_{t+1})^{\rm T} = \zeta^{\rm{H}}_{t+1}\bf{\mathcal{V}}_{t+1}[\bf{I}_{(t+1)\times t} \;\; \zeta_{t+1}]  
\EE
with the error covariance matrix of $\{ \bf{x}_1,...,\bf{x}_{t}, \phi_t(\bf{r}_t) \}$ defined as
\BE\label{Eqn:V_phi_a}
 \bf{\mathcal{V}}_{t+1} \equiv \left[
\begin{array}{cc}
        \bf{V}_{t}^{\bar{\phi}}   &  \begin{array}{c}
            v^{{\phi}}_{1,t+1}  \\
            \vdots  \end{array}  \\ 
        \begin{array}{cc}
            v^{{\phi}}_{t+1, 1} &    \cdots 
        \end{array}  & v^{{\phi}}_{t+1,t+1}
\end{array}
\right]_{(t+1) \times (t+1)},
\EE
where $\bf{v}^{\bar{\phi}}_{t+1} = [v^{\bar{\phi}}_{t+1,1},...,v^{\bar{\phi}}_{t+1,t+1}]^{\rm T}$, $v^{\phi}_{t+1,t^{'}}\equiv\tfrac{1}{N}\mr{E}\{[\phi_t(\bf{r}_t)-\bf{x}]^{\rm{H}}\bf{f}_{t^{'}}\}$, and the corresponding covariance vector $\bf{v}^{\phi}_{t+1} = [v^{\phi}_{t+1,1},...,v^{\phi}_{t+1,t+1}]^{\rm T}$ can be calculated by the Monte Carlo method (see details in~\cite[Appendix C and Appendix H]{MAMPTIT}).

\emph{Note:} The SE holds for MAMP receiver under the assumptions of Lipschitz-continuous $\hat{\gamma}_t(\cdot)$ and $\hat{\phi}_t(\cdot)$, which correspond to linear constraint and code constraint in coded GMIMO, respectively. Since  $\hat{\gamma}_t(\cdot)$ in \eqref{Eqn:localLMMF} and LDPC decoder have been proved to be  Lipschitz-continuous respectively in \cite{MAMPTIT} and \cite[Appendix B]{LC-LDPC}, the SE holds for MAMP receiver with LDPC decoding $\hat{\phi}_t(\cdot)$. Therefore, a kind of LDPC code is designed for MAMP receiver in simulation results. Although there is no strict proof for other types of FEC codes, we conjecture that $\hat{\phi}_t(\cdot)$ is also Lipschitz-continuous for the majority of FEC codes (e.g., Turbo code, Polar code, Reed-solomon (RS) code, etc.).

\section{Coding Principle and Information-Theoretic Optimality of MAMP}
In this section, we present the achievable rate analysis, optimal coding principle, and information-theoretic optimality proof of MAMP. The practical LDPC code optimization for MAMP is also given for ill-conditioned and correlated channel matrices, respectively.

\subsection{Achievable Rate Analysis and Coding Principle}\label{coding principle}
To circumvent the complex multi-dimensional SE analysis of MAMP, we first prove the fixed-point consistency of MAMP and OAMP/VAMP as follows, based on the Bayes optimality of MAMP in uncoded systems.

\begin{lemma}[Fixed-Point Consistency]\label{lem:same_fp}
Let the SE fixed point of MAMP in \eqref{Eqn:SE} be $(v_{*}^{\gamma}, v_{*}^{\phi})$, where $v_{*}^{\gamma}=\lim\limits_{t \to \infty } v_{t,t}^{\gamma}$ and $v_{*}^{\phi}=\lim\limits_{t \to \infty } v_{t,t}^{\phi}$. MAMP and OAMP/VAMP have the same SE fixed point $(v_{*}^{\gamma}, v_{*}^{\phi})$ for arbitrary fixed Lipschitz-continuous $\hat{\phi}_t(\cdot)$ in NLD.
\end{lemma}
\begin{IEEEproof}
In uncoded systems, the Bayes optimality of MAMP is proved in \cite[Theorem 2]{MAMPTIT}, i.e., MAMP and OAMP/VAMP can converge to the same SE fixed point $(v_{*}^{\gamma}, v_{*}^{\phi})$. Note that there is no restriction on the specific form of $\hat{\phi}_t(\cdot)$ in the proof of \cite[Theorem 2]{MAMPTIT}. Therefore, for arbitrary fixed Lipschitz-continuous $\hat{\phi}_t(\cdot)$, MAMP and OAMP/VAMP have the same SE fixed point $(v_{*}^{\gamma}, v_{*}^{\phi})$.
\end{IEEEproof}

Based on Lemma~\ref{lem:same_fp}, the multi-dimensional SE of MAMP can converge to the same SE fixed point as the SISO SE of OAMP/VAMP for the same APP decoder $\hat{\phi}_t(\cdot)$. This inspires us to attempt to analyze the achievable rate of MAMP with the aid of the SISO SE of OAMP/VAMP.

\begin{figure*}[!t]
    \centering
    \subfigure[Equivalent MAMP receiver: $\eta_t(\cdot)$ denotes the enhanced MLD consisting of $\gamma_t(\cdot)$, damping and orthogonal operations. $\hat{\phi}_t(\cdot)$ denotes the demodulation and APP decoder.]
    {
        \begin{minipage}[t]{0.7\linewidth}\label{Fig:MAMP receiver1} 
            \centering
            \includegraphics[width=0.8\linewidth]{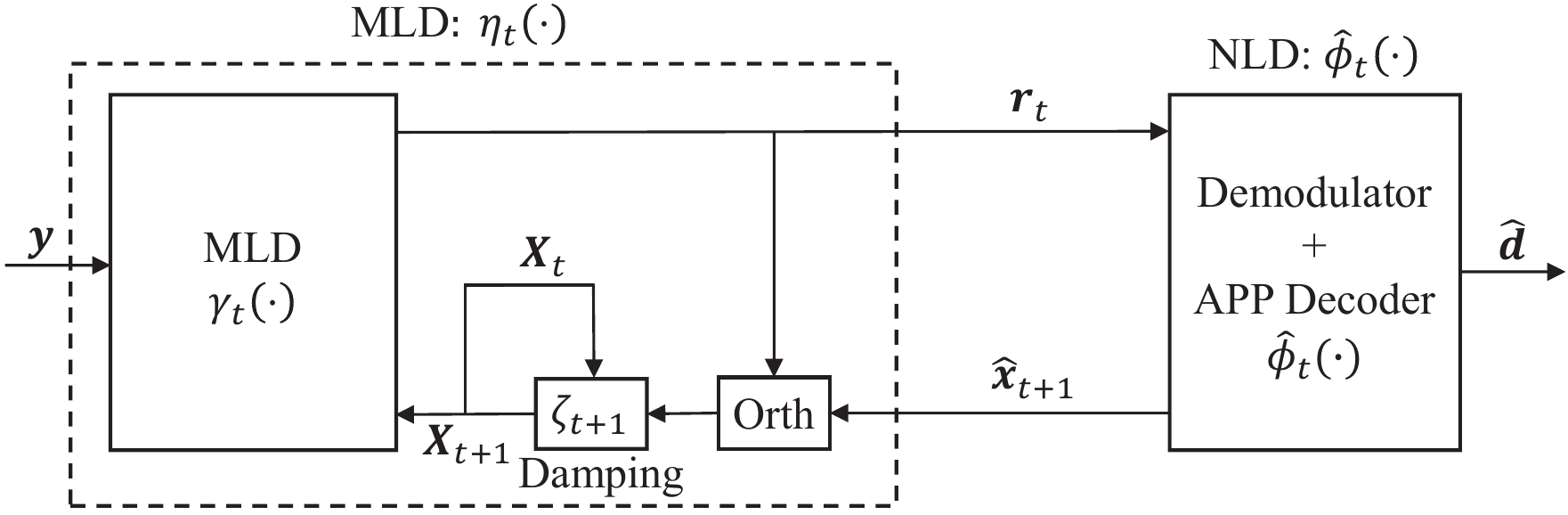}\vspace{0.3cm}
		\end{minipage}
	}\\
	\subfigure[Variational transfer functions: $\eta_{\rm SE}$ and $\hat{\phi}_{\rm SE}$ are the variational transfer functions of $\eta_t$ and $\hat{\phi}_t$, respectively.]
	{
		\begin{minipage}[t]{0.7\linewidth}\label{Fig:MAMP SE1} 
			\centering
			\includegraphics[width=0.75\linewidth]{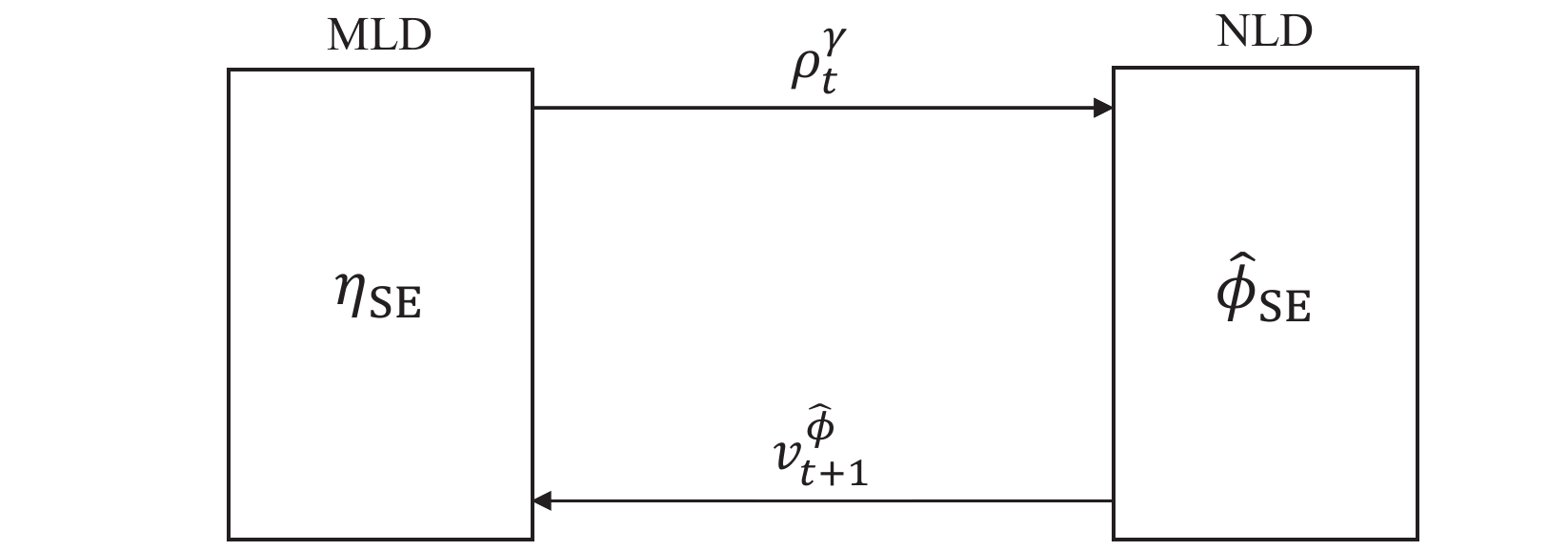}\vspace{0.3cm}
		\end{minipage} 
	}
	\caption{Illustration of the equivalent MAMP receiver and its variational state evolution.}\vspace{0.2cm}
\end{figure*}

Specifically, an equivalent transformation of MAMP in \eqref{Eqn:BO-MAMP} is obtained by incorporating all orthogonal and damping operations into the MLD, as shown in Fig.~\ref{Fig:MAMP receiver1}, in which the equivalent MAMP is given by
\BS\label{Eqn:BO-MAMP1}
\begin{align}
\mathrm{MLD}: \quad \;\;\;\; \bf{r}_t = \eta_t(\hat{\bf{x}}_t), \\
\mathrm{NLD}: \quad  \hat{\bf{x}}_{t+1} = \hat{\phi}_t(\bf{r}_t),
\end{align}
\ES
where $\eta_t(\cdot)$ is a multi-dimensional MLD involving $\gamma_t(\cdot)$ in \eqref{Eqn:MAMP-MLE}, damping, and orthogonal operations, and $\hat{\bf{x}}_t$ denotes the output \emph{a posteriori} estimation of APP decoder $\hat{\phi}_t(\cdot)$.  It is worth noting that this equivalent transformation does not change the SE fixed point (i.e., convergence performance) of MAMP. As a result, the equivalent MAMP is also referred to as MAMP for simplicity. Importantly, in contrast to the complexly multi-dimensional transfer function in \eqref{Eqn:SE-NLE}, the transfer function of $\hat{\phi}_t(\cdot)$ in NLD is SISO. However, since the memory is required in $\eta_t(\cdot)$, the transfer function of $\eta_t(\cdot)$ 
remains intricately multi-dimensional. This continues to impede the theoretical analysis of MAMP.

To overcome the above issue, a SISO variational transfer function $\eta_{\rm{SE}}$ of MLD $\eta_t(\cdot)$ is derived with the aid of Lemma~\ref{lem:same_fp} and the SE of OAMP/VAMP. Therefore, as shown in Fig.~\ref{Fig:MAMP SE1}, a SISO variational SE (VSE) of MAMP is presented in the following lemma, which is adopted to simplify the achievable rate analysis and optimal code design for MAMP.

\begin{lemma}[VSE of MAMP]\label{lem:VSE_MAMP}
Let $\rho_t^{\gamma} = 1/ v_{t,t}^{\gamma}$ and $v^{\hat{\phi}}_{t}\equiv\tfrac{1}{N}\mr{E}\{\left \| \hat{\bf{x}}_t -\bf{x} \right \|^2\}$ denote the signal-to-interference-plus-noise ratio (SINR) of $\bf{r}_t$ and the MSE of $\hat{\bf{x}}_t$, respectively. The VSE transfer functions of MAMP per transmit antenna can be written as
\BS\label{Eqn:VSE_MAMP}
\begin{align}
    \mathrm{MLD}:\;\; \;\; \rho^{\gamma}_{t} &= \eta_{\rm{SE}}(v^{\hat{\phi}}_t) = (v^{\hat{\phi}}_t)^{-1}-[\hat{\gamma}_{\mr{SE}}^{-1}(v^{\hat{\phi}}_t)]^{-1} ,\label{Eqn:VSE-MLD}\\
    \mathrm{NLD}:\; v^{\hat{\phi}}_{t+1} &= \hat{\phi}^{\mathcal{C}}_{\rm{SE}}(\rho^{\gamma}_{t}) = \mr{mmse} \{\bf{x}|\sqrt{\rho^{\gamma}_{t}}\bf{x}+\bf{z}, \Phi_{\mathcal{C}}\},\label{Eqn:VSE-NLD}
\end{align}
\ES
where $\hat{\gamma}_{\mr{SE}}(v)= \tfrac{1}{N}{\rm{tr}}\{[snr\bf{A}^{\rm{H}}\bf{A}+v^{-1}\bf{I}]^{-1}\}$ denotes the MSE function of LMMSE detector, $\hat{\gamma}_{\mr{SE}}^{-1}(\cdot)$ the inverse of $\hat{\gamma}_{\mr{SE}}(\cdot)$, and $\bf{z}\sim \mathcal{CN}(\bf{0}, \bf{I})$ an AWGN vector independent of $\bf{x}$.
\end{lemma}
\begin{IEEEproof}
    See Appendix~\ref{APP:VSE_MAMP}.
\end{IEEEproof}

Note that the VSE transfer functions in~\eqref{Eqn:VSE_MAMP} are not equivalent to the SE transfer functions in~\eqref{Eqn:SE}. Although VSE cannot be utilized to characterize the MSE performance of MAMP in each iteration, it can be employed to accurately analyze the achievable rate and coding principle.

Due to the coding gain, the decoding transfer function $\hat{\phi}_{\mr{SE}}^{\mathcal{C}}(\cdot)$ is upper bounded by the demodulation transfer function $\hat{\phi}_{\mr{SE}}^{\mathcal{S}}(\cdot)$, i.e.,
\BE\label{Eqn:CodeMAMP1}
\hat{\phi}^{\mathcal{C}}_{\rm{SE}}(\rho^{\gamma}_{t}) < \hat{\phi}^{\mathcal{S}}_{\rm{SE}}(\rho^{\gamma}_{t}),  \quad \mr{for}\;\; 0\leq \rho^{\gamma}_{t} \leq \rho_{\rm{max}},	
\EE
where $\hat{\phi}^{\mathcal{S}}_{\rm{SE}}(\rho^{\gamma}_{t}) = \mr{mmse} \{\bf{x}|\sqrt{\rho^{\gamma}_{t}}\bf{x}+\bf{z}, \Phi_{\mathcal{S}}\}$ and $\rho_{\rm{max}} = \mathcal{J} \cdot snr$ with \footnote{For $\beta \ge 1$, $\frac{1}{N}{\rm tr}\{\bf{A}^{\rm{H}}\bf{A}\}=1$ and  $\tfrac{1}{N}\mr{E}\{\|\bf{y}\|^2\}=1$, such that $\rho_{\rm{max}}=snr$.  When $\beta<1$, $\frac{1}{M}{\rm tr}\{\bf{A}^{\rm{H}}\bf{A}\}=1$ and  $\tfrac{1}{N}\mr{E}\{\|\bf{y}\|^2\}=\tfrac{M}{N}$, so $\rho_{\rm{max}}=\tfrac{snr}{\beta}$.}
\BE
    \mathcal{J} = 
    \left\{ \begin{aligned}
	  & 1 ,  \;\;\;\;\;\;\;\; \beta \ge 1\\
	  & 1/\beta , \;\;\;\;  \beta < 1
    \end{aligned} 
    \right. .
\EE

\begin{figure}[!tbp]
	\centering
	\includegraphics[width=0.78\columnwidth]{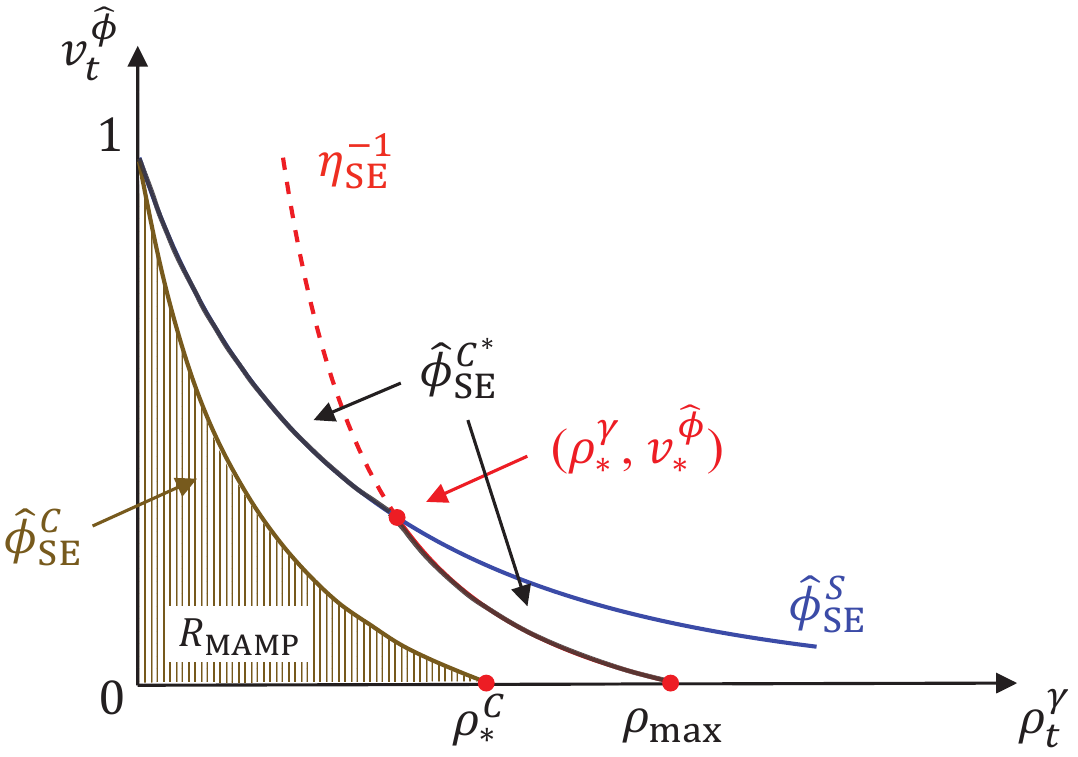}
	\caption{Graphical illustration of VSE for MAMP, where $\eta^{-1}_{\rm{SE}}(\cdot)$ is the inverse function of $\eta_{\rm{SE}}(\cdot)$. $\hat{\phi}^{\mathcal{S}}_{\rm{SE}}(\cdot)$ and $\hat{\phi}^{\mathcal{C}}_{\rm{SE}}(\cdot)$ with $\hat{\phi}_{\mr{SE}}^{\mathcal{C}}(\rho_{*}^{\mathcal{C}})=0$ denote the MMSE functions of constellation and code constraint in NLD, respectively. $(\rho_*^{\gamma}, v_*^{\hat{\phi}})$ denotes the VSE fixed point between $\eta^{-1}_{\rm{SE}}(\cdot)$ and $\hat{\phi}^{\mathcal{S}}_{\rm{SE}}(\cdot)$. Moreover, $\hat{\phi}^{\mathcal{C}^*}_{\rm{SE}}(\cdot)$ is the optimal coding function of MAMP.}\label{Fig:SE curve}
\end{figure}

As shown in Fig.~\ref{Fig:SE curve}, assume that there is a unique fixed point $(\rho_*^{\gamma}, v_*^{\hat{\phi}})$ between $\eta_{\rm{SE}}^{-1}(\cdot)$ and $\hat{\phi}^{\mathcal{S}}_{\rm{SE}}(\cdot)$. Since $v_*^{\hat{\phi}}>0$, the converge performance of MAMP is not error-free. Therefore, to achieve the error-free performance, a kind of proper FEC code should be well-designed to guarantee an available decoding tunnel between $\eta_{\rm{SE}}^{-1}(\cdot)$ and $\hat{\phi}^{\mathcal{C}}_{\rm{SE}}(\cdot)$. That is, 
\BE\label{Eqn:CodeMAMP2}
    \hat{\phi}^{\mathcal{C}}_{\rm{SE}}(\rho^{\gamma}_{t}) < \eta_{\rm{SE}}^{-1}(\rho^{\gamma}_{t}), \quad \mr{for}\;\; 0 \leq \rho^{\gamma}_{t} \leq \rho_{\rm{max}}.
\EE

Therefore, based on \eqref{Eqn:CodeMAMP1} and \eqref{Eqn:CodeMAMP2}, we obtain the error-free condition of MAMP in the following lemma.
\begin{lemma}[Error-Free Decoding]\label{lem:err_free}
MAMP can achieve error-free decoding if and only if 
\BE\label{Eqn:CodeMAMP3}
\!\!\!\hat{\phi}^{\mathcal{C}}_{\rm{SE}}(\rho^{\gamma}_{t}) <  \min\{\hat{\phi}^{\mathcal{S}}_{\rm{SE}}(\rho^{\gamma}_{t}), \eta_{\rm{SE}}^{-1}(\rho^{\gamma}_{t})\},  \; \mr{for}\; 0\leq \rho^{\gamma}_{t} \leq \rho_{\mr{max}}.
\EE
\end{lemma}

Then, based on Lemma~\ref{lem:err_free} and I-MMSE lemma\cite{GuoTIT2005}, we give the achievable rate of MAMP as follows.

\begin{lemma}[Achievable Rate of MAMP]\label{lem:Rate}
The achievable rate of MAMP per transmit antenna with fixed $\hat{\phi}_{\mr{SE}}^{\mathcal{C}}(\cdot)$ is
\BE \label{Eqn:rate_MAMP}
\begin{aligned}
 &R_{\text{MAMP}} = \int_{0}^{\rho_{*}^{\mathcal{C}}}
 \hat{\phi}_{\mr{SE}}^{\mathcal{C}}(\rho_t^{\gamma}) d \rho_t^{\gamma},   \\
 &\begin{array}{l@{\quad}l}
 {\rm s.t.} &  \hat{\phi}_{\mr{SE}}^{\mathcal{C}}(\rho_t^{\gamma})< \hat{\phi}_{\mr{SE}}^{\mathcal{C}^*}(\rho_t^{\gamma}), \;\;  \mr{for}\;\;\;0\le \rho_t^{\gamma} \le \rho_{\mr{max}},
 \end{array}
\end{aligned}
\EE    
where $\rho_{*}^{\mathcal{C}}=\hat{\phi}_{\mr{SE}}^{\mathcal{C}^{-1}}(0)$ and $\hat{\phi}_{\mr{SE}}^{\mathcal{C}^*}(\rho_t^{\gamma})=\mr{min}\{\hat{\phi}^{\mathcal{S}}_{\rm{SE}}(\rho_t^{\gamma}), \eta_{\rm{SE}}^{-1}(\rho_t^{\gamma})\}$.
\end{lemma}

Therefore, based on Lemma~\ref{lem:Rate}, the optimal code design principle of MAMP can be obtained in the following lemma.

\begin{lemma}[Optimal Code Design]\label{lem:optimal code design}
    The optimal coding principle of MAMP is
    \BE\label{Eqn: Optimal coded desdign}
    \hat{\phi}_{\mr{SE}}^{\mathcal{C}}(\rho_t^{\gamma}) \to \hat{\phi}_{\mr{SE}}^{\mathcal{C}^*}(\rho_t^{\gamma}), \quad \mr{for}\;\;\; 0\le \rho_t^{\gamma} \le \rho_{\mr{max}}, 
    \EE
    enabling MAMP to achieve error-free performance as well as the maximum achievable rate.
\end{lemma}

Based on Lemma~\ref{lem:optimal code design}, the maximum achievable rate of MAMP is obtained directly in the following theorem.

\begin{theorem}[Maximum Achievable Rate]\label{The:maxrate}
The maximum achievable rate of MAMP per transmit antenna is 
\BE\label{Eqn:rateMAMP}
  	R_{\rm{MAMP}}^{\rm{max}} \to  \int_{0}^{\rho_{\rm{max}}}  \hat{\phi}^{\mathcal{C}^*}_{\rm{SE}}(\rho_t^{\gamma}) d\rho_t^{\gamma},
\EE
where $\hat{\phi}^{\mathcal{C}^*}_{\rm{SE}}(\rho_t^{\gamma})={\rm{min}}\{\hat{\phi}^{\mathcal{S}}_{\rm{SE}}(\rho_t^{\gamma}), \eta_{\rm{SE}}^{-1}(\rho_t^{\gamma})\}$.
\end{theorem}

\begin{figure*}[!t]
	\centering
	\subfigure[Maximum achievable rate comparison between MAMP and CAS-MAMP for GMIMO with $\bf{A}_{\rm ill}$, where $N=500$, $\kappa=10$, and $\beta=\{0.67,1.00,1.50\}$.]
	{
		\begin{minipage}[t]{0.75\linewidth}\label{Fig:AR for ill channel} 
			\centering
			\includegraphics[width=1\linewidth]{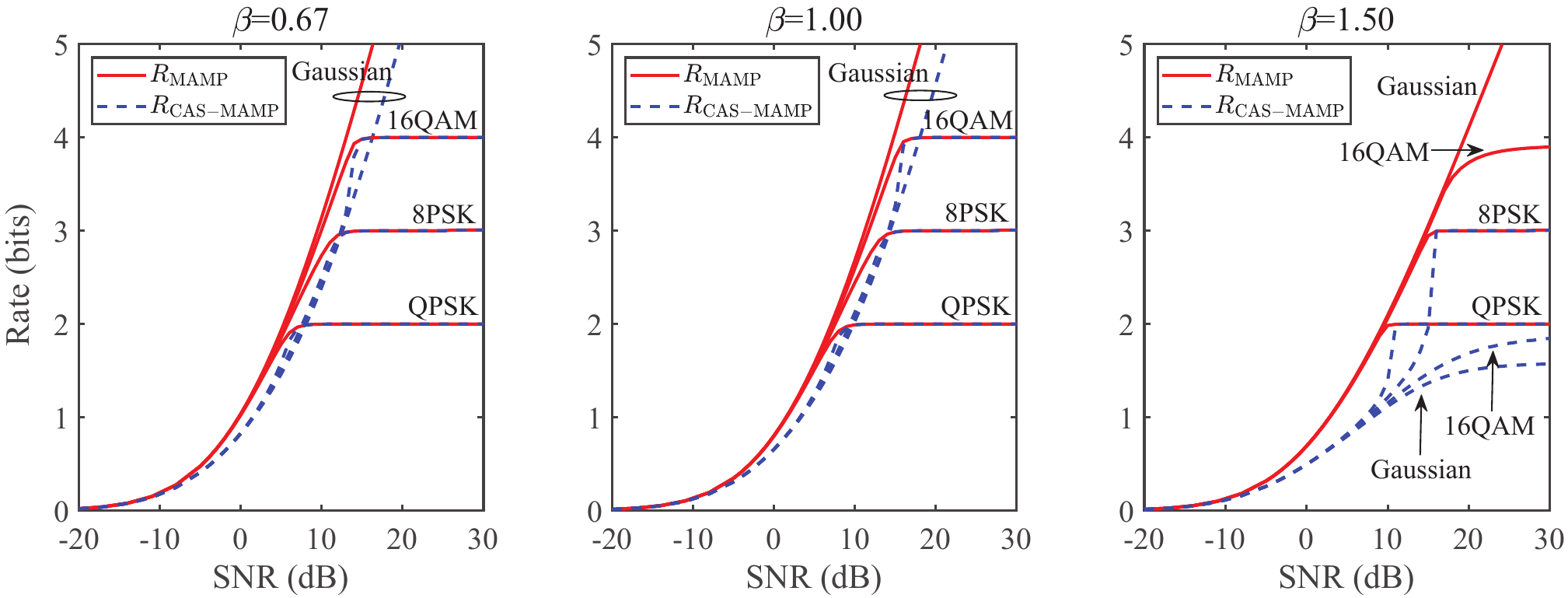}\vspace{0.2cm}
		\end{minipage}
	}\\
	\subfigure[Maximum achievable rate comparison between MAMP and CAS-MAMP for GMIMO with $\bf{A}_{\rm cor}$, where $N=500$, $\beta=1.0$, and $\alpha=\{0.2,0.4,0.6\}$.]
	{
		\begin{minipage}[t]{0.75\linewidth}\label{Fig:AR for correlated channel} 
			\centering
			\includegraphics[width=1\linewidth]{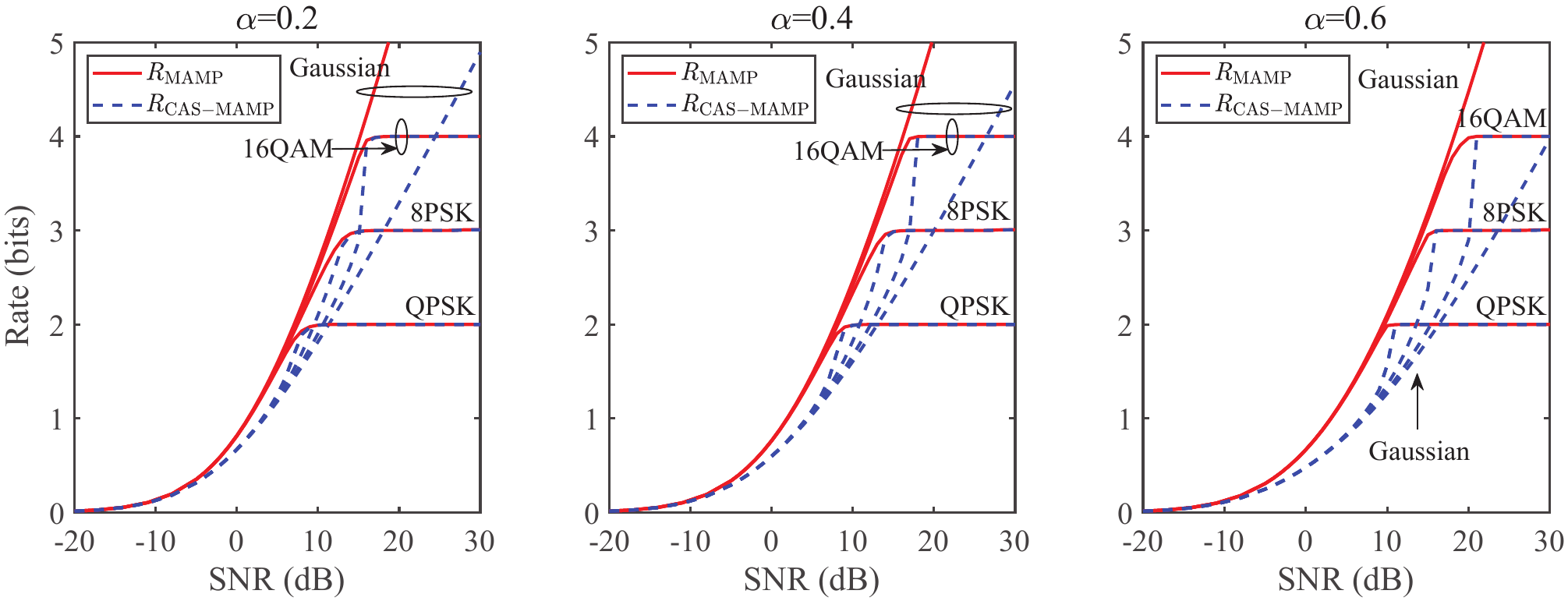}\vspace{0.2cm}
		\end{minipage} 
	}
	\caption{Maximum achievable rate comparison between MAMP and CAS-MAMP for coded GMIMO with $\bf{A}_{\rm ill}$ and $\bf{A}_{\rm cor}$, where QPSK, 8PSK, 16QAM and Gaussian signaling are considered.}\label{Fig:AR} 
\end{figure*}

\subsection{Information-Theoretic Optimality of MAMP}
Due to the constrained-capacity optimality of OAMP/VAMP~\cite{LeiOptOAMP,YuhaoTcom2022}, based on Lemma~\ref{lem:VSE_MAMP}, Lemma~\ref{lem:optimal code design}, and Theorem~\ref{The:maxrate}, the information-theoretic (i.e., constrained-capacity) optimality of MAMP is verified in the following theorem.
\begin{theorem}[Constrained-Capacity Optimality]\label{The:Constrained-Capacity Optimality}
MAMP can achieve the same maximum achievable rate as OAMP/VAMP, which indicates MAMP is constrained-capacity optimal in GMIMO, i.e.,
\BS\label{Eqn:capcacity}
\begin{align}
	   &R_{\rm{MAMP}}^{\rm{max}} = R_{\rm{OAMP/VAMP}}^{\rm{max}}\to \int_{0}^{\rho_{\rm{max}}}  \hat{\phi}^{\mathcal{C}^*}_{\rm{SE}}(\rho) d\rho,   \\
         {\rm s.t.} &  \quad \hat{\phi}^{\mathcal{C}}_{\rm{SE}}(\rho) \to \hat{\phi}^{\mathcal{C}^*}_{\rm{SE}}(\rho) = \rm{min}\{ \hat{\phi}^{\mathcal{S}}_{\rm{SE}}(\rho), \eta^{-1}_{\rm{SE}}(\rho)\},
\end{align}
\ES
where $R_{\rm{MAMP}}^{\rm{max}}$ is equal to the average constrained capacity of GMIMO per transmit antenna given in \cite{LeiOptOAMP,YuhaoTcom2022}. As a result, the maximum achievable sum rate of MAMP is
\BE
R_{\rm{MAMP}}^{\rm{sum}}=NR_{\rm{MAMP}}^{\rm{max}},
\EE
which is the same as the constrained sum capacity of GMIMO.
\end{theorem}

Furthermore, and most importantly, we derive a more general theorem inspired by Lemma~\ref{lem:same_fp}. This theorem states that any two iterative detection algorithms with the same SE fixed point can be analyzed equivalently from the perspective of theory, even if these two algorithms have different local detectors and SE transfer functions of different dimensions. Therefore, Theorem~\ref{The:Constrained-Capacity Optimality} can be viewed as a special case of this more general theorem.

\begin{theorem}[Theoretical-Analysis Equivalence]\label{The:capacity_optimal}
If two iterative detection algorithms have the same SE fixed point, then their achievable rate analyses, optimal coding principles, and maximum achievable rates are all the same.
\end{theorem}
\begin{IEEEproof}
We consider arbitrary iterative detection algorithm $1$ and algorithm $2$ have the same SE fixed point, such that algorithm $1$ and algorithm $2$ can achieve the same converged performances. Assume the optimal coding principle of algorithm $1$ as 
\BE\label{Eqn:CodeMAMP4}
\hat{\phi}^{\mathcal{C}}_{\rm{SE}}(\rho) <  \hat{\phi}^{\mathcal{C}^*_1}_{\rm{SE}}(\rho),  \quad \mr{for}\;\; 0\leq \rho \leq \rho_{\mr{max}},
\EE
where $\hat{\phi}^{\mathcal{C}^*_1}_{\rm{SE}}(0)=1$ and $\hat{\phi}^{\mathcal{C}^*_1}_{\rm{SE}}(\rho_{\rm max})=0$. Therefore, algorithm~$2$ can also achieve the error-free performance as algorithm~$1$ when given $\hat{\phi}^{\mathcal{C}}_{\rm{SE}}(\rho)<\hat{\phi}^{\mathcal{C}^*_1}_{\rm{SE}}(\rho)$ for $\rho\in[0,\rho_{\rm max}]$. This is the sufficient condition of error-free decoding for algorithm 2. Next, the converse method is used to show that it is also the necessary condition of error-free decoding for algorithm~$2$.

Given $\forall \rho_1,\rho_2 \in[0,\rho_{\rm max}]$ with $\rho_1<\rho_2$. Assume that algorithm~$2$ can achieve error-free performance when $\hat{\phi}^{\mathcal{C}}_{\rm{SE}}(\rho)\ge\hat{\phi}^{\mathcal{C}^*_1}_{\rm{SE}}(\rho)$ for $\rho\in[\rho_1,\rho_2]$ and $\hat{\phi}^{\mathcal{C}}_{\rm{SE}}(\rho)<\hat{\phi}^{\mathcal{C}^*_1}_{\rm{SE}}(\rho)$ for $\rho\in[0,\rho_1]\cup[\rho_2,\rho_{\rm max}]$. Then, algorithm~$1$ can also achieve error-free performance with the above condition, which contradicts \eqref{Eqn:CodeMAMP4}.

As a result, \eqref{Eqn:CodeMAMP4} is the necessary and sufficient condition of error-free decoding for algorithm~$2$. Based on I-MMSE lemma~\cite{GuoTIT2005}, achievable rates of algorithms~$1$ and $2$ are the same, i.e.,
\BE 
     R_2 = R_1 = \int_{0}^{\hat{\phi}_{\mr{SE}}^{\mathcal{C}^{-1}}(0)} \hat{\phi}_{\mr{SE}}^{\mathcal{C}}(\rho) d \rho, 
\EE  
where $\hat{\phi}_{\mr{SE}}^{\mathcal{C}}(\rho) < \hat{\phi}^{\mathcal{C}^*_1}_{\rm{SE}}(\rho)$ for $0\le \rho \le \rho_{\mr{max}}$. Therefore, the optimal coding principle of algorithm~$2$ is the same as that of algorithm $1$, which enables algorithms~$2$ and $1$ to achieve error-free performance with the same maximum achievable rate as follows:
\BE
     R_{2}^{\rm{max}} = R_{1}^{\rm{max}}\to \int_{0}^{\rho_{\rm{max}}}  \hat{\phi}^{\mathcal{C}^*_1}_{\rm{SE}}(\rho) d\rho, 
\EE

Therefore, we complete the proof of Theorem~\ref{The:capacity_optimal}.
\end{IEEEproof}

Based on Lemma~\ref{lem:same_fp} and Theorem~\ref{The:capacity_optimal},  the constrained-capacity optimality of MAMP can be proved, i.e., MAMP and OAMP/VAMP have the identical maximum achievable rate, which is equal to the constrained capacity of GMIMO.

\begin{figure*}[!t]
	\centering
	\includegraphics[width=0.9\linewidth]{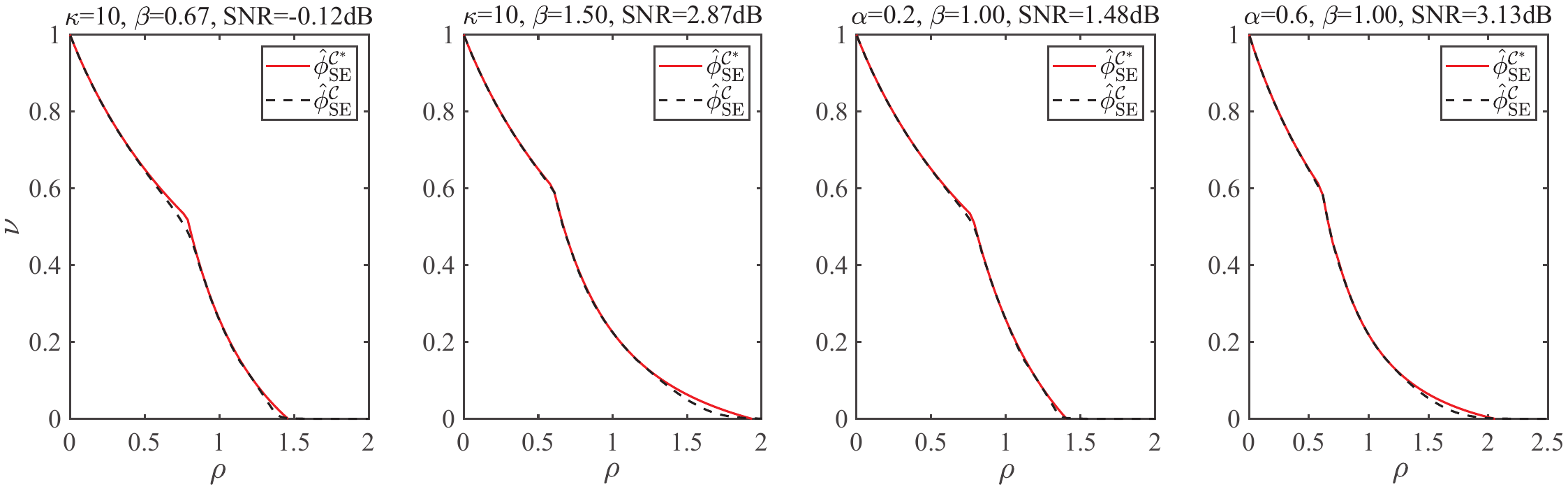}
	\caption{The SE curves of desired $\{\hat{\phi}^{\mathcal{C}^{*}}_{\rm{SE}}\}$ and designed $\{\hat{\phi}^{\mathcal{C}}_{\rm{SE}}\}$ for MAMP in GMIMO with $\bf{A}_{\rm ill}$ and $\bf{A}_{\rm cor}$, where $N=500$, $\beta\in\{0.67, 1.00, 1.50\}$, $\kappa=10$, and $\alpha \in \{0.2, 0.6\}$.}\label{Fig:curve match}\vspace{0.2cm}
\end{figure*}

\begin{table*}[!htbp]
	\centering 
	\caption{Optimized LDPC Codes for MAMP in GMIMO with Ill-Conditioned, Rayleigh Fading, and Correlated Channel Matrices}
	\scalebox{0.6}{
		\begin{tabular}{|m{2.7cm}<{\centering}||m{2.5cm}<{\centering}|m{2.5cm}<{\centering}|m{2.5cm}<{\centering}|m{2.5cm}<{\centering}|m{2.5cm}<{\centering}|m{2.5cm}<{\centering}|m{2.5cm}<{\centering}|m{2.5cm}<{\centering}|m{2.5cm}<{\centering}|} \hline
			
			Channel type &  \multicolumn{6}{c|}{$\bf{A}_{\rm ill}$} & \multicolumn{1}{c|}{$\bf{A}_{\rm Ray}$} & \multicolumn{2}{c|}{ $\bf{A}_{\rm cor}$}   \\ \hline
			$\beta $ &  \multicolumn{2}{c|}{0.67} &  \multicolumn{2}{c|}{1.00} &  \multicolumn{2}{c|}{1.50}  
			&  \multicolumn{3}{c|}{1.00}  \\ \hline
			$\kappa/ \alpha $ & $\kappa=10$ & $\kappa=50$ & $\kappa=10$ & $\kappa=50$ & $\kappa=10$ & $\kappa=50$ & $\alpha=0$ & $\alpha=0.2$ & $\alpha=0.6$ \\ \hline
			$N$ & \multicolumn{9}{c|}{500}\\ \hline
			$M$ &  \multicolumn{2}{c|}{750} &  \multicolumn{2}{c|}{500} &  \multicolumn{2}{c|}{333}   &  \multicolumn{3}{c|}{500}\\ \hline
			Codeword length & \multicolumn{9}{c|}{$1\times10^5$}   \\ \hline
			$R_{\rm{LDPC}}$ & 0.5050 & 0.5049 & 0.5094 & 0.5062 & 0.5059 & 0.4983 & 0.5058 & 0.5063 & 0.5061   \\ \hline
			$R_{\rm{sum}}$ & 505.0 & 504.9 & 509.4 & 506.2 & 505.9 & 498.3 & 505.8 & 506.3 & 506.1   \\ \hline
			$\mu(X)$ & \multicolumn{2}{c|}{$\mu_7=1$} & \multicolumn{3}{c|}{$\mu_8=1$} & $\mu_8=0.8$ $\mu_{30}=0.2$ & \multicolumn{3}{c|}{$\mu_8=1$} \\ \hline
			$\lambda(X)$ & $\lambda_2=0.3649$ $\lambda_3=0.2353$ $\lambda_9=0.1741$ $\lambda_{27}=0.2257$  
			& $\lambda_2=0.5348$ $\lambda_{16}=0.2654$  $\lambda_{17}=0.0532$   $\lambda_{100}=0.1195$ $\lambda_{110}=0.0271$ 
			& $\lambda_2=0.3143$ $\lambda_3=0.2094$  $\lambda_9=0.1236$   $\lambda_{10}=0.0700$ $\lambda_{30}=0.0858$ $\lambda_{35}=0.1109$  $\lambda_{80}=0.0860$
			& $\lambda_2=0.4623$ $\lambda_3=0.0021$  $\lambda_{14}=0.2510$   $\lambda_{15}=0.0129$ $\lambda_{70}=0.1086$ $\lambda_{80}=0.0685$  $\lambda_{900}=0.0947$
			& $\lambda_2=0.4222$ $\lambda_3=0.0690$ $\lambda_{15}=0.0139$ $\lambda_{16}=0.2474$ $\lambda_{70}=0.1355$ $\lambda_{200}=0.1088$ $\lambda_{800}=0.0032$
			& $\lambda_2=0.3840$ $\lambda_{16}=0.1511$ $\lambda_{17}=0.1560$ $\lambda_{90}=0.1592$ $\lambda_{800}=0.1497$
			& $\lambda_2=0.2950$ $\lambda_3=0.2235$ $\lambda_8=0.0769$ $\lambda_9=0.1285$ $\lambda_{30}=0.0790$ $\lambda_{35}=0.1108$ $\lambda_{70}=0.0863$
			& $\lambda_2=0.3040$ $\lambda_3=0.2180$ $\lambda_8=0.0028$ $\lambda_9=0.1917$ $\lambda_{30}=0.0280$ $\lambda_{35}=0.1789$ $\lambda_{90}=0.0766$
			& $\lambda_2=0.4544$ $\lambda_3=0.0034$ $\lambda_{11}=0.0835$ $\lambda_{12}=0.1556$ $\lambda_{50}=0.0099$ $\lambda_{60}=0.2342$ $\lambda_{800}=0.0591$  \\ \hline
			$(\rm{SNR})^{\ast}_{\text{dB}}$   & -0.12 & 1.48 & 1.64 & 3.20 & 2.87 & 5.35 & 1.33 & 1.48 & 3.13 \\ \hline
			${\text{(Capacity)}_{\text{dB}}}$ & -0.19 & 1.40 & 1.55 & 3.15 & 2.85 & 5.03 & 1.30 & 1.45 & 3.10 \\ \hline
	\end{tabular}}\label{table:parameters1} 
\end{table*}

\subsection{Maximum Achievable Rates of MAMP in Ill-Conditioned and Correlated Channels}\label{sec:MAR}
The proposed achievable rate analysis of MAMP is available for general right-unitarily-invariant channel matrices. Here, we provide the maximum achievable rates of MAMP using the common ill-conditioned channel matrices $\bf{A}_{\rm{ill}}$ and correlated channel matrices $\bf{A}_{\rm cor}$ as examples, which are widely used in practical communication scenario and modelled as follows:
\subsubsection{Ill-conditioned channel matrix}
Let the SVD of $\bf{A}_{\rm ill}$ be $\bf{A}_{\rm ill} = \bf{U}\bf{\Lambda}\bf{V}^{\rm{H}}$. $\bf{U}$ and $\bf{V}$ are generated by unitary matrices of SVD decomposition of an IID Gaussian matrix. We set the eigenvalues $\{e_i\}$ in $\bf{\Lambda}$ as\cite{Vila2015ICASSP}: $e_i/e_{i+1} = \kappa^{1/\mathcal{L}}, i=1,...,\mathcal{L}-1$, and $\sum\nolimits_{i = 1}^{\mathcal{L}} {e_i^2 = \mathcal{F}}$, where $\kappa \ge 1$ denotes the condition number of $\bf{A}_{\rm ill}$, $\mathcal{L}\!=\!{\rm{min}}\{M,N\}$, and $\mathcal{F}={\rm {max}} \{M, N\}$.
\subsubsection{Correlated channel matrix}
The correlated channel matrix $\bf{A}_{\rm cor}$ is constructed using
the Kronecker channel model~\cite{00Da}: $\bf{A}_{\rm cor}=\bf{C}_{R}^{\frac{1}{2}}{\tilde{\bf{A}}}\bf{C}_{T}^{\frac{1}{2}}$, where $\tilde{\bf{A}} \in \mathbb{C}^{M\times N}$ is an IID Gaussian matrix, and $\bf{C}_{R} \in \mathbb{R}^{M\times M}$ and $\bf{C}_{T} \in \mathbb{R}^{N\times N}$ are the receive and transmit correlation matrices, respectively. Moreover, the $(i,j)$-th element of both $\bf{C}_{R}$ and $\bf{C}_{T}$ is $\alpha^{|i-j|}$, where $\alpha \in [0,1)$ denotes the correlation coefficient and a larger $\alpha$ correlates to a stronger antenna correlation. 
It is worth noting that when $\alpha=0$, the correlated channel matrix $\bf{A}_{\rm cor}$ is degraded to the IID Gaussian matrix $\tilde{\bf{A}}$, i.e., the classical and commonly used Rayleigh fading channel matrix $\bf{A}_{\rm Ray}=\tilde{\bf{A}}$~\cite{Ray1}.
For simplicity, the correlation coefficients for $\bf{C}_{R}$ and $\bf{C}_{T}$ are set to be the same.

Based on Theorem~\ref{The:Constrained-Capacity Optimality}, Fig.~\ref{Fig:AR} illustrates the maximum achievable rates of MAMP in coded GMIMO with $N=500$ and different signaling (i.e., QPSK, 8PSK, 16QAM, and Gaussian signaling). In Fig.~\ref{Fig:AR for ill channel}, it is shown that the maximum achievable rates of MAMP increase with the modulation order and SNR for $\bf{A}_{\rm ill}$ with $\kappa=10$ and $\beta=\{0.67,1.00,1.50\}$. The corresponding maximum achievable rates for QPSK, 8PSK, and 16QAM converge to a constant. As shown in Fig.~\ref{Fig:AR for correlated channel}, the maximum achievable rates of MAMP with $\bf{A}_{\rm cor}$ are similar to that with $\bf{A}_{\rm ill}$, where $\beta=1.0$ and $\alpha=\{0.2,0.4,0.6\}$.

To demonstrate the advantages of MAMP, we also present the maximum achievable rates of the conventional CAS-MAMP receiver~\cite{CDMA2002,CDMA2005}, in which the MLD and NLD are implemented sequentially without iteration over each other. Based on I-MMSE lemma\cite{GuoTIT2005}, Fig.~\ref{Fig:SE curve} shows that the maximum achievable rate of CAS-MAMP is $R_{\rm{CAS-MAMP}}^{\text{max}} = \int_{0}^{\rho^{\gamma}_{*}}  \hat{\phi}^{\mathcal{C}^*}_{\rm{SE}}(\rho_t^{\gamma}) d\rho_t^{\gamma}$ for given SNR.
As a result, compared with $R^{\text{max}}_{\rm MAMP}$, the rate loss of CAS-MAMP is $ R_{\rm{loss}}= \int_{\rho^{\gamma}_{*}}^{\rho^{\text{max}}}  \hat{\phi}^{\mathcal{C}^*}_{\rm{SE}}(\rho_t^{\gamma}) d\rho_t^{\gamma}$. As shown in Fig.~\ref{Fig:AR}, MAMP can obtain higher achievable rates than CAS-MAMP with $\bf{A}_{\rm ill}$ and $\bf{A}_{\rm cor}$. Moreover, compared to MAMP, the achievable rate of CAS-MAMP with high-order modulation (e.g., Gaussian signaling and 16QAM) is lower than that of low-order modulation (e.g., QPSK) in harsh scenarios, where $\beta=1.5$ or $\alpha=0.6$ (similar phenomena has been discussed in \cite{LeiTIT2021}).

\subsection{Practical Code Design for MAMP}
Considering that LDPC code is widely used in wireless communication~\cite{ryan2009channel} and LDPC decoder is proved to be Lipschitz-continuous in \cite[Appendix B]{LC-LDPC}, we propose a kind of LDPC code $\mathcal{C}(\lambda(X), \mu(X))$ for MAMP, where $\lambda(X)=\textstyle\sum_{d = 2}^{d_{v,\rm{max}}} {{\lambda_{d}}{X^{d- 1}}}$ and $\mu(X) = \textstyle\sum_{d = 2}^{d_{c,\rm{max}}} {{\mu_{d}}{X^{d - 1}}}$ are the degree distributions of variable and check nodes, respectively, and $d_{v,\rm{max}}$ and $d_{c,\rm{max}}$ are the corresponding maximum degrees of variable and check nodes. Based on Lemma~\ref{lem:optimal code design}, LDPC codes are designed for ill-conditioned and correlated channel matrices with different parameters. The details about the optimization of LDPC codes can refer to~\cite[Section IV-A]{LeiTIT2021}.

With QPSK modulation, the optimized LDPC codes for MAMP with $\bf{A}_{\rm ill}$ and $\bf{A}_{\rm cor}$ are given in Table~\ref{table:parameters1}, in which $R_{\rm{LDPC}}$ denotes the optimized LDPC code rate and the sum rate is $R_{\rm{sum}}=NR_{\rm{LDPC}}\log_2|{\cal{S}}_{\rm{QPSK}}|$ with $|{\cal{S}}_{\rm{QPSK}}|=4$. In coded GMIMO, the gaps between the theoretical decoding thresholds $(\rm{SNR})^{\ast}_{\text{dB}}$ of the optimized LDPC codes and the associated constrained capacities per transmit antenna are within about 0.3 dB. Meanwhile, Fig.~\ref{Fig:curve match} shows that for $\bf{A}_{\rm ill}$ and $\bf{A}_{\rm cor}$, the SE curves of designed $\{\hat{\phi}^{\mathcal{C}}_{\rm{SE}}\}$ with optimized LDPC codes match well with those of desired $\{\hat{\phi}^{\mathcal{C}^*}_{\rm{SE}}\}$ in~\eqref{Eqn: Optimal coded desdign}, which also illustrates the optimality of the optimized LDPC codes for MAMP.

\begin{figure*}[!t]\vspace{0.1cm}
    \centering
    \subfigure[$\kappa=10$]{\includegraphics[width=0.4\linewidth]{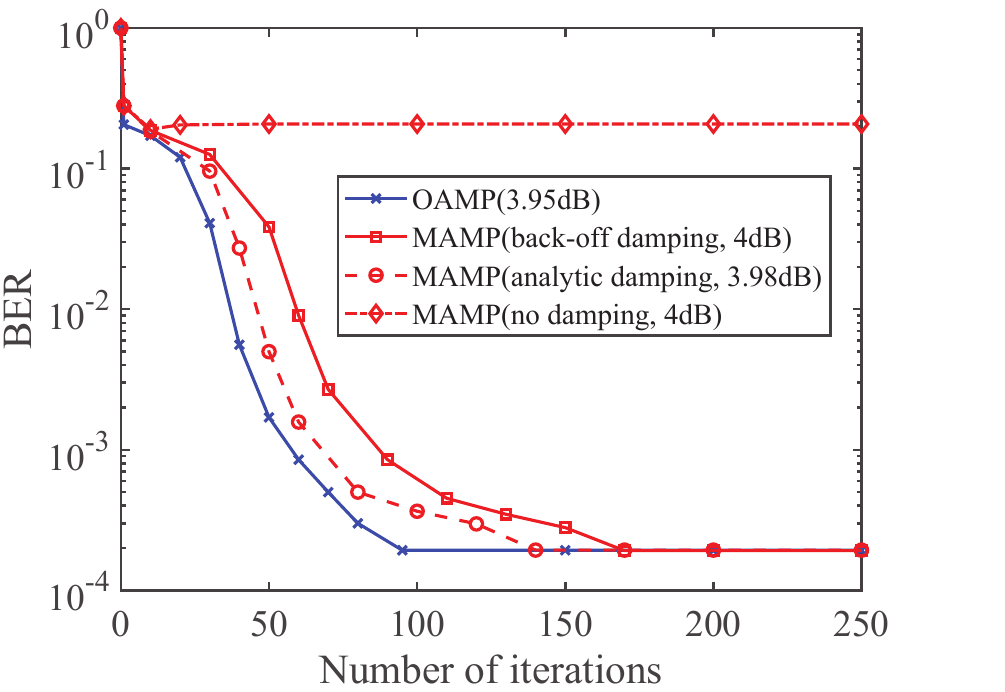}\label{Fig:damping10}}
    \subfigure[$\kappa=50$]{\includegraphics[width=0.4\linewidth]{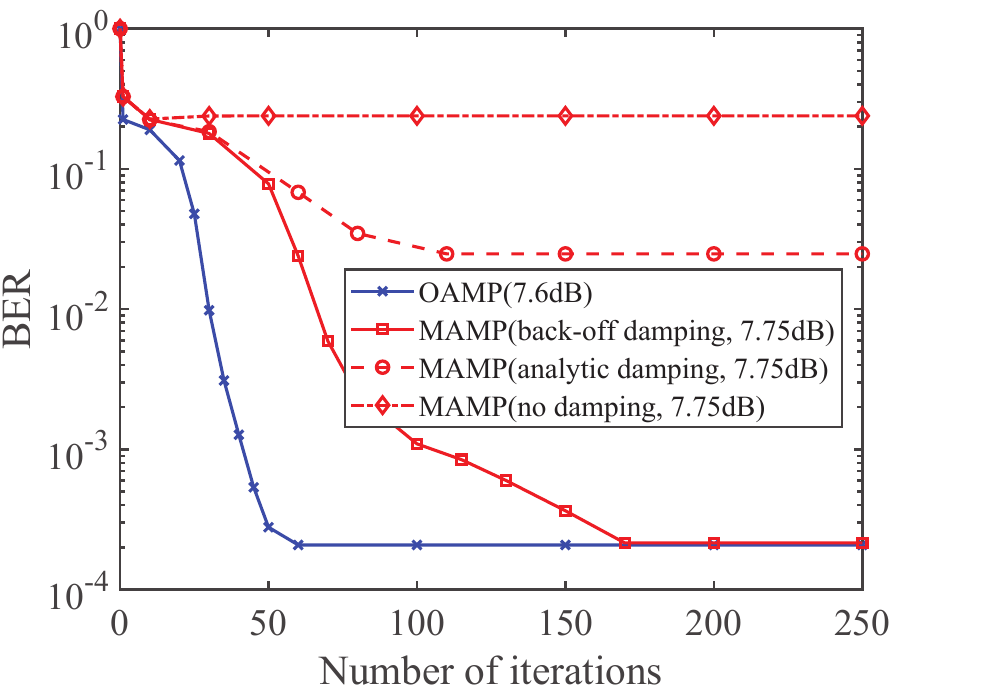}\label{Fig:damping50}}
	
    \caption{Convergence comparison of MAMP and OAMP/VAMP with analytical and back-off damping, where target ${\text{BER}}=2\times 10^{-4}$, $\beta=1.5$, $(M,N)=(333,500)$, and the optimized LDPC codes are given in Table~\ref{table:parameters1}.}\label{Fig:damping}\vspace{0.1cm}
\end{figure*}

\section{Numerical Results}
In this section, we present the BER performances and running time complexity of MAMP with optimized LDPC codes in GMIMO. Meanwhile, BER performance comparisons with existing schemes are provided.

\vspace{-0.2cm}
\subsection{Simulation Configuration}
Note that the optimal coding principle and information-theoretic optimality of MAMP is available for arbitrary input distributions.  In the following simulations, the practical finite-length LDPC codes are designed for QPSK. However, for high-order modulations (e.g., 16QAM), it is difficult to optimize practical LDPC codes by traditional code design methods such as EXIT analysis directly due to the asymmetry of the bits in higher-order modulation constellation points~\cite{YuhaoTcom2022}. As a result, the code design for high-order modulations is left as our future work. In addition, ill-conditioned channel matrix $\bf{A}_{\rm ill}$ and correlated channel matrix $\bf{A}_{\rm cor}$ are set up in Section~\ref{sec:MAR}.

\subsection{Effect of Damping on MAMP Convergence}
To validate the effect of damping on MAMP, we present the convergence comparison of MAMP with the analytical damping and back-off damping in Fig.~\ref{Fig:damping}, with target $\text{BER}=2\times 10^{-4}$, ill-conditioned channel matrix $\bf{A}_{\rm ill}$, $\kappa=\{10, 50\}$, $\beta=1.5$, $N=500$, and optimized LDPC codes in Table~\ref{table:parameters1}. The two types of damping methods for MAMP are defined as follows.

\begin{enumerate}
\item \emph{Analytical damping~\cite[Lemma 8]{MAMPTIT}:}
\BE\label{Eqn:a-damping}
\bf{\zeta}^{\mathcal{I}}_{t+1}=\left\{ \begin{aligned}
	&\frac{ \left [ \bf{\mathcal{V}}^{\mathcal{I}}_{t+1} \right ]^{-1} \bf{1}}{\bf{1}^{\rm T}\left [ \bf{\mathcal{V}}^{\mathcal{I}}_{t+1} \right ]^{-1} \bf{1}} ,  \;\;\; \rm{if}\; \bf{\mathcal{V}}^{\mathcal{I}}_{t+1} \; \rm{is\; invertible}\\
	&\left [ 0,\cdots\!,1,0  \right ]^{\rm T} ,  \;\;\;\;\; \rm{otherwise} 
    \end{aligned} 
    \right. ,
\EE
where $\bf{\zeta}^{\mathcal{I}}_{t+1}$ is the optimized damping vector for $\phi_t(\bf{r}_t)$ and non-trivial memories $\{\bf{x}_i,i\in\mathcal{I}_t\}$, $\bf{\mathcal{V}}^{\mathcal{I}}_{t+1}$ the corresponding covariance matrix, and $\mathcal{I}_t$ the index set of non-trivial memories for damping. The damping length is set to $L_d=|\mathcal{I}_t|+1=3$ in the following simulations.

\item \emph{Back-off damping:}
\BE\label{Eqn:b-damping}
\bf{\zeta}_{t+1}=\left\{ \begin{aligned}
	\left [ 0,\cdots\!,1,0  \right ]^{\rm T} ,  \;\;\; v^{\phi}_{t+1,t+1}>v^{\phi}_{t,t}\\
	\left [ 0,\cdots\!,0,1  \right ]^{\rm T} , \;\;\;  v^{\phi}_{t+1,t+1}\le v^{\phi}_{t,t}
    \end{aligned} 
    \right. .
\EE
This indicates that when the output variance of $\phi_t(\cdot)$ increases, the estimated output value and variance from the previous iteration are taken as the current output of $\bar{\phi}_t(\cdot)$, which is utilized to prevent MAMP from diverging during the iteration.
\end{enumerate}

In Fig.~\ref{Fig:damping}, OAMP/VAMP with optimized LDPC codes in Table~\ref{table:parameters1} is employed as the optimal bound, and MAMP without damping is used as the baseline. Note that OAMP/VAMP converges to the target $\text{BER}=2\times 10^{-4}$ with minimum iterations for $\kappa=10$ and $50$, but MAMP without damping (i.e., no damping in Fig.~\ref{Fig:damping}) converges to the BER as high as $2\times 10^{-1}$. As shown in Fig.~\ref{Fig:damping10}, although MAMP with analytical damping converges to OAMP/VAMP for $\kappa=10$, it only achieves $\text{BER}=2\times 10^{-2}$ for $\kappa=50$ in Fig.~\ref{Fig:damping50}. In contrast, whether for $\kappa=10$ or $50$, MAMP with back-off damping can always converge to OAMP/VAMP. The main reason is that the decoding tunnel between the transfer functions of MLD and NLD is extremely narrow for MAMP with optimized LDPC codes, as illustrated in Fig.~\ref{Fig:curve match}, which results in the  element values of $\bf{\mathcal{V}}^{\mathcal{I}}_t$ in~\eqref{Eqn:a-damping} being quite near, causing $\bf{\mathcal{V}}^{\mathcal{I}}_t$ to be close to the singular matrix. This leads to instability and poor performance of MAMP with analytical damping due to inverse of $\bf{\mathcal{V}}^{\mathcal{I}}_t$ in~\eqref{Eqn:a-damping}. As a result, the back-off damping is employed for MAMP in the following simulations.

\begin{figure*}[!tbp]

	\centering
	\subfigure[$\kappa=10$ and $\beta=\{0.67,1.00,1.50\}$]{\includegraphics[width=0.95\linewidth]{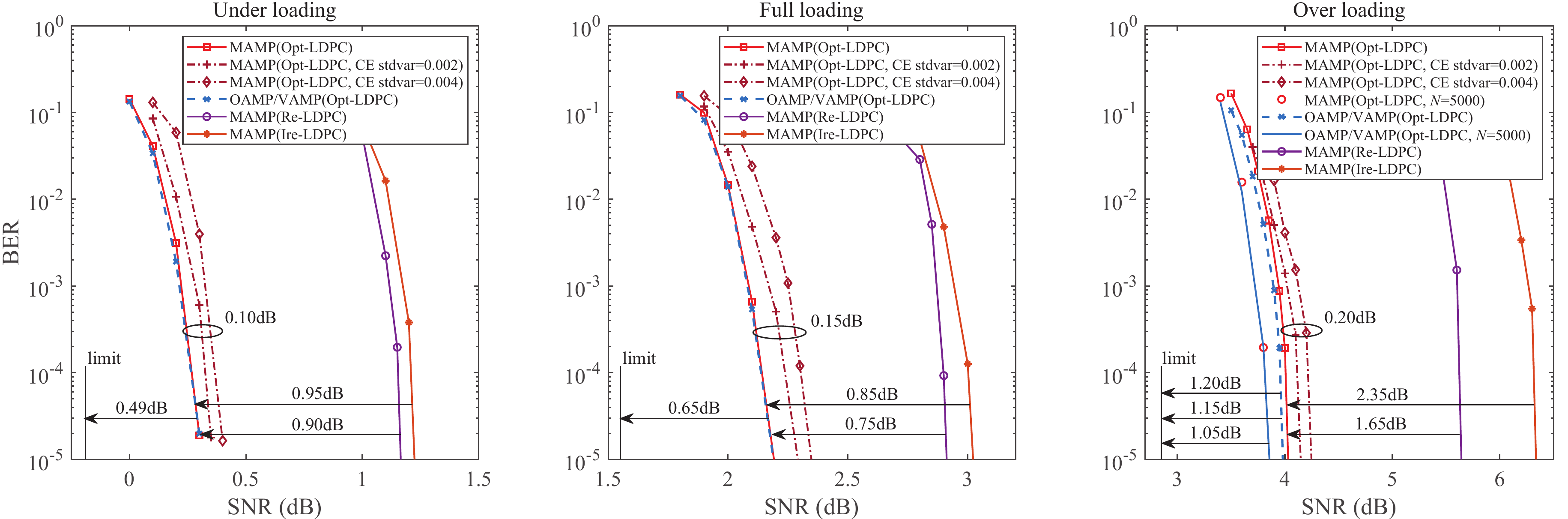}\label{Fig:BER10}}\\
	\subfigure[$\kappa=50$ and $\beta=\{0.67,1.00,1.50\}$]{\includegraphics[width=0.95\linewidth]{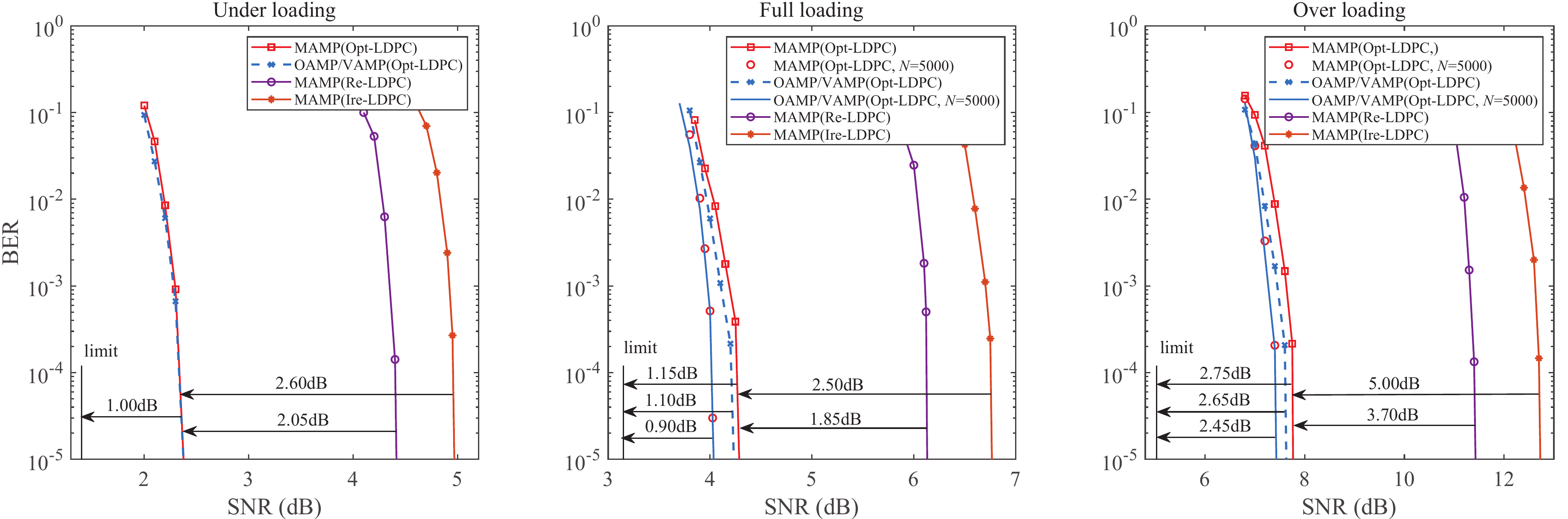}}
	
	\caption{BER performances of MAMP and OAMP/VAMP with optimized LDPC codes and MAMP with P2P LDPC codes for $\bf{A}_{\rm ill}$, where $\kappa=\{10,50\}$, $N=\{500,5000\}$, and code length = $\{1\times10^5,2\times10^5\}$. Underloading, full-loading, and overloading correspond to $\beta=0.67$, $\beta=1$, and $\beta=1.5$, respectively. ``Opt-LDPC'' is the optimized LDPC code in Table~\ref{table:parameters1} and Table~\ref{table:parameters2}, ``Re-LDPC'' the P2P-regular (3,6) LDPC code with $R_{\rm{LDPC}}=0.5$~\cite{ryan2009channel}, ``Ire-LDPC'' the P2P-irregular code with $R_{\rm{LDPC}}=0.5$~\cite{Richardson2001}, ``limit'' the associated constrained capacity, and ``CE stdvar''  the standard deviations of channel estimation errors. Except for specific labeling, $N$ is equal to $500$.}\label{Fig:BER} 
\end{figure*}

\subsection{BER Performance of MAMP for Ill-Conditioned Channel Matrices}
\subsubsection{BER comparison to OAMP/VAMP with optimized LDPC codes}
As shown in Fig.~\ref{Fig:BER}, the BER comparisons between MAMP and OAMP/VAMP with optimized LDPC codes in Table~\ref{table:parameters1} are presented, where $\kappa=\{10,50\}$, $\beta=\{0.67,1.00,1.50\}$, and $N=\{500,5000\}$. Note that MAMP can achieve the same BER performances as OAMP/VAMP for ($\beta=\{0.67,1.00\}$, $\kappa=10$) and  ($\beta=0.67$, $\kappa=50$). As $\beta$ and $\kappa$ increase, i.e., ($\beta=1.5$, $\kappa=\{10,50\}$) and ($\beta=1$, $\kappa=50$), the gaps between BER curves at $10^{-5}$ of MAMP and OAMP/VAMP are still within $0.1$ dB, since the optimality analysis for MAMP is based on infinite length assumption.
Meanwhile, we provide the BER comparisons between MAMP and OAMP/VAMP for large-scale systems, i.e., $N=5000$, in which the corresponding optimized LDPC codes are given in Table~\ref{table:parameters2}. It is worth noting that MAMP can achieve the same BER performances as OAMP/VAMP in large-scale systems, but with much lower complexity than OAMP/VAMP.

\begin{table}[!tbp] 
	\centering 
	\caption{Optimized LDPC Codes for MAMP in Large-Scale Systems}
	\scalebox{0.55}{        
		\begin{tabular}{|m{3cm}<{\centering}||m{3.5cm}<{\centering}|m{3.5cm}<{\centering}|m{3.5cm}<{\centering}|} \hline
			
			Channel type &  \multicolumn{3}{c|}{ill-conditioned channel matrices} \\ \hline
			$\beta $ &  1.00 &  \multicolumn{2}{c|}{1.50}  \\ \hline
			$\kappa$ & 50 & 10 &50 \\ \hline
			$N$ & \multicolumn{3}{c|}{5000}\\ \hline
			$M$ &  5000  &  \multicolumn{2}{c|}{3333}  \\ \hline
			Codeword length & \multicolumn{3}{c|}{$2\times10^5$}   \\ \hline
			$R_{\rm{LDPC}}$ & 0.5062 & 0.5059 & 0.4985    \\ \hline
			$R_{\rm{sum}}$  & 5062.0  & 5059.0 & 4985.0   \\ \hline
			$\mu(X)$ & \multicolumn{2}{c|}{$\mu_8=1$} & $\mu_8=0.8 , \mu_{30}=0.2$   \\ \hline
			$\lambda(X)$ & $\lambda_2=0.4623$ $\lambda_3=0.0021$ $\lambda_{14}=0.2511$ $\lambda_{15}=0.0128$ $\lambda_{70}=0.1087$ $\lambda_{80}=0.0685$ $\lambda_{900}=0.0947$
			& $\lambda_2=0.4222$ $\lambda_3=0.0687$  $\lambda_{15}=0.1149$   $\lambda_{16}=0.1383$ $\lambda_{70}=0.1680$ $\lambda_{350}=0.0879$
			& $\lambda_2=0.3842$ $\lambda_{16}=0.1589$ $\lambda_{17}=0.1475$ $\lambda_{90}=0.1640$ $\lambda_{900}=0.1454$ \\ \hline
			$(\rm{SNR})^{\ast}_{\text{dB}}$   & 3.20 & 2.87 & 5.35  \\ \hline
			${\text{(Capacity)}_{\text{dB}}}$ & 3.15 & 2.85 & 5.03  \\ \hline
	\end{tabular}}\label{table:parameters2} 
\end{table}

\subsubsection{BER comparison with P2P regular and irregular LDPC codes}
To validate the advantages of the optimized LDPC codes, we also present the BER performances of MAMP with P2P regular and well-designed irregular LDPC codes in Fig.~\ref{Fig:BER}. The parameters of the P2P-regular LDPC codes are $(3, 6)$ LDPC code with coding rate $R_{\rm{LDPC}}=0.5$ \cite{ryan2009channel}. The degree distributions of the well-designed P2P-irregular LDPC code are $\lambda(X)=0.24426x+0.25907x^2+0.01054x^3+0.05510x^4+0.01455x^7+0.01275x^9+0.40373x^{11}$ and $\mu(X)=0.25475x^6+0.73438x^7+0.01087x^8$, whose rate $R_{\rm{LDPC}}$ is $0.5$ and the decoding threshold is 0.18~dB away from the P2P-AWGN capacity~\cite{Richardson2001}. As shown in Fig.~\ref{Fig:BER}, the gaps between BER curves at $10^{-5}$ of the optimized LDPC codes and the associated constrained capacities are $0.49 \sim 2.75$ dB, which verifies the capacity-approaching performances of the optimized LDPC codes. Moreover,  MAMP with the optimized codes have $0.75\sim 5$ dB gains over the MAMP with P2P-regular and well-designed P2P-irregular LDPC codes, where $\beta=\{0.67, 1, 1.5\}$ and $\kappa=\{10, 50\}$. This also indicates the Bayes-optimal MAMP with P2P regular and well-designed irregular LDPC codes are not optimal anymore with  significant performance losses in GMIMO.

\begin{figure}[!htbp]
	\centering 
	\includegraphics[width=0.92\columnwidth]{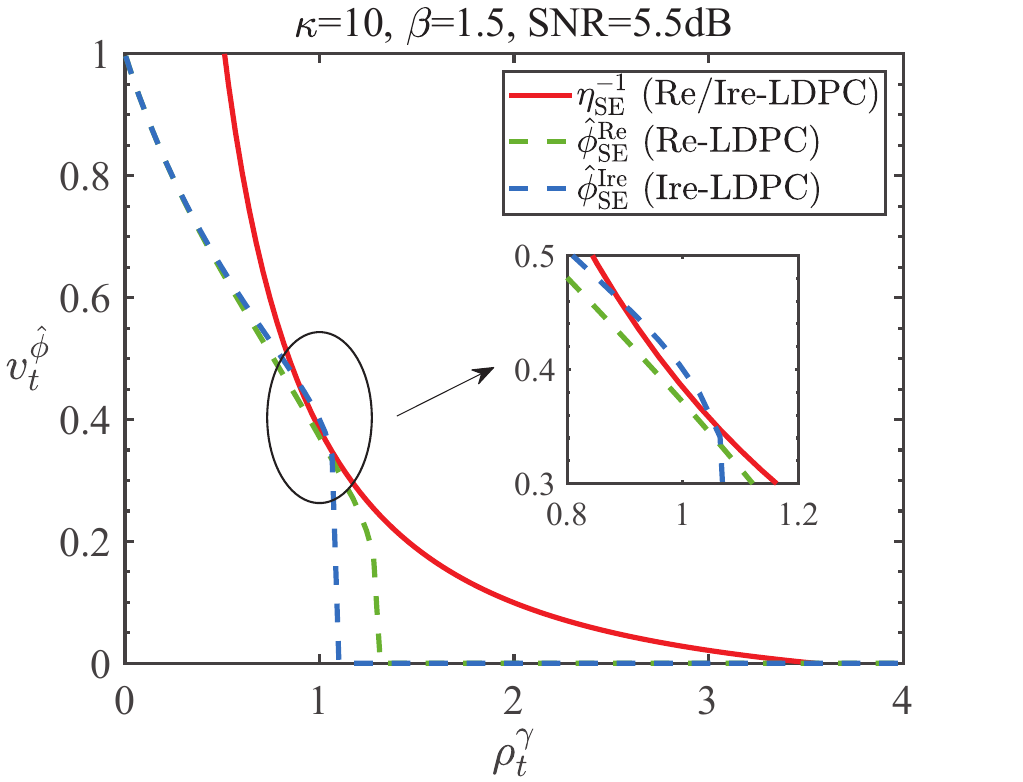} 
	\caption{The VSE of MAMP for $\bf{A}_{\rm ill}$, where $\kappa=10$, $\beta=1.5$, and $N=500$. $\eta^{-1}_{\rm{SE}}(\cdot)$ is the inverse function of $\eta_{\rm{SE}}(\cdot)$. $\hat{\phi}^{\rm{Re}}_{\rm{SE}}(\cdot)$ and $\hat{\phi}^{\rm{Ire}}_{\rm{SE}}(\cdot)$ denote the MMSE function of the code constraint corresponding to P2P-regular (3,6) LDPC code (``Re-LDPC'')~\cite{ryan2009channel} and P2P-irregular LDPC code (``Ire-LDPC'')~\cite{Richardson2001}, respectively.} \label{Fig:Re/Ire-LDPC} 
\end{figure}

In addition, as shown in Fig.~\ref{Fig:BER}, the BER performances of MAMP with P2P-regular (3,6) LDPC code are superior to those of MAMP with P2P-irregular LDPC code. This is determined by the iterative process between the MLD and NLD of MAMP, which can be accurately predicted by VSE. Therefore, taking $\bf{A}_{\rm ill}$ with $\kappa=10$, $\beta=1.5$, and $N=500$ as an example, Fig.~\ref{Fig:Re/Ire-LDPC} shows the VSE transfer curves of MAMP with P2P regular and irregular LDPC codes, where $\eta_{\rm{SE}}^{-1}(\cdot)$ is the inverse function of the MLD transfer function $\eta_{\rm{SE}}(\cdot)$, $\hat{\phi}^{\rm{Re}}_{\rm{SE}}(\cdot)$ and $\hat{\phi}^{\rm{Ire}}_{\rm{SE}}(\cdot)$ denote the NLD transfer functions corresponding to P2P-regular (3,6) LDPC code and P2P-irregular LDPC code, respectively. Note that for P2P-regular (3,6) LDPC code, there is an available decoding tunnel between $\eta_{\rm{SE}}^{-1}(\cdot)$ and $\hat{\phi}^{\rm{Re}}_{\rm{SE}}(\cdot)$ to achieve the error-free performance. 
In contrast, for P2P-irregular LDPC code, $\eta_{\rm{SE}}^{-1}(\cdot)$ and $\hat{\phi}^{\rm{Ire}}_{\rm{SE}}(\cdot)$ intersect prematurely. Therefore, the decoding process of P2P-irregular LDPC code will be terminated in advance, making it impossible to achieve the error-free performance.

\subsubsection{BER performance with imperfect channel state information}
To confirm the robustness of MAMP with optimized codes, we consider the BER simulations in GMIMO with imperfect channel estimations, where $\kappa=10$, $N=500$, $\beta=\{0.67,1.00,1.50\}$, and the standard deviations of estimated channel errors are 0.002 and 0.004. As shown in Fig.~\ref{Fig:BER10}, the imperfect channel estimations cause about $0.05 \sim 0.2$ dB performance losses, in which the gaps between BER curves at $10^{-5}$ of the optimized LDPC codes and the associated constrained capacities are within 1.4 dB. This verifies that MAMP with optimized codes are robust to the imperfect channel estimations.

 \begin{figure*}[!t]
	
	\centering
	\includegraphics[width=0.95\linewidth]{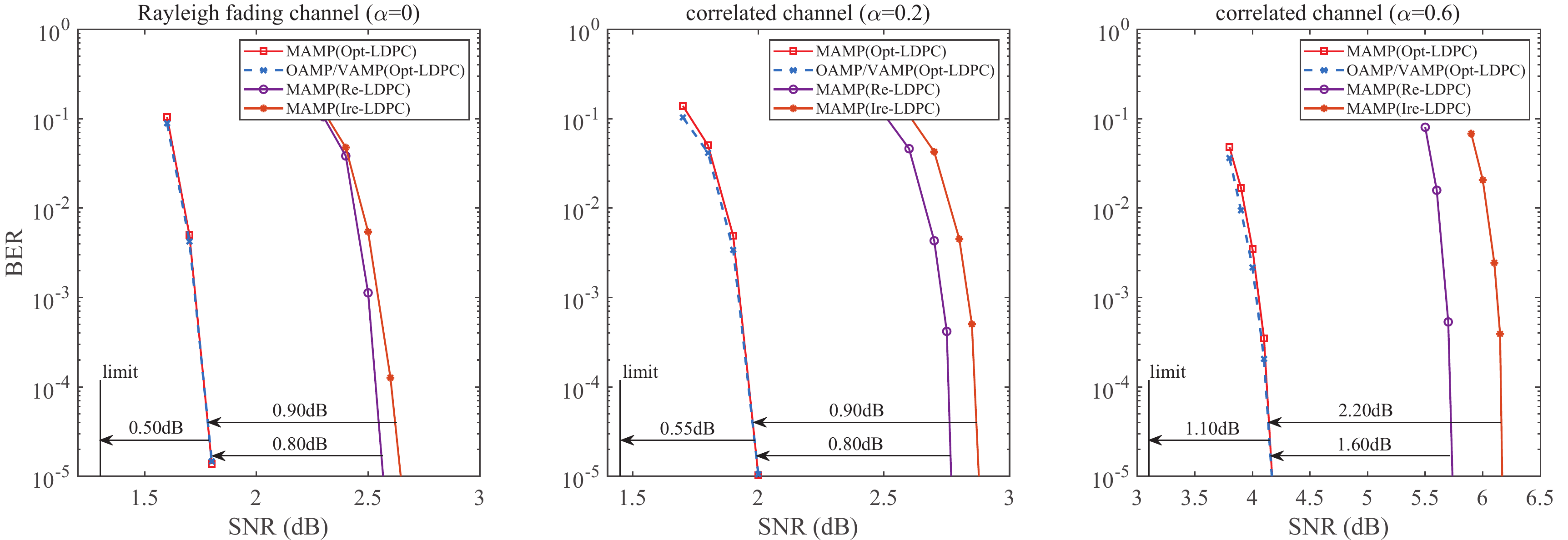} 
	
	\caption{BER performances of MAMP and OAMP/VAMP with optimized LDPC codes and MAMP with P2P LDPC codes for $\bf{A}_{\rm Ray}$ and $\bf{A}_{\rm cor}$, where $\alpha=\{0,0.2,0.6\}$, $M=N=500$, and code length = $10^5$. ``Opt-LDPC'' is the optimized LDPC code in Table~\ref{table:parameters1}, ``Re-LDPC'' the P2P-regular (3,6) LDPC code with $R_{\rm{LDPC}}=0.5$~\cite{ryan2009channel}, ``Ire-LDPC'' the P2P-irregular code with $R_{\rm{LDPC}}=0.5$~\cite{Richardson2001}, and ``limit'' the associated constrained capacity.}\label{Fig:correlated_BER} \vspace{0.5cm}
\end{figure*}

\subsection{BER Performance of MAMP for Correlated Channel Matrices}
Fig.~\ref{Fig:correlated_BER} shows the BER performances of MAMP and OAMP/VAMP in GMIMO with $\bf{A}_{\rm cor}$, where $N=500$, $\beta=1$, $\alpha=\{0,0.2,0.6\}$, and the optimized LDPC codes are given in Table \ref{table:parameters1}. 
Particularly, $\bf{A}_{\rm cor}$ is degraded to $\bf{A}_{\rm Ray}$ when $\alpha=0$. 
As shown in Fig.~\ref{Fig:correlated_BER}, MAMP and OAMP/VAMP with optimized LDPC codes have the same BER performances. Furthermore, the BER curves at $10^{-5}$ of MAMP are $0.5\sim1.1$ dB away form the associated constrained capacities, which confirms the capacity-approaching performances of the optimized LDPC codes for $\bf{A}_{\rm Ray}$ and $\bf{A}_{\rm cor}$. In addition, MAMP with optimized codes have $0.8\sim 2.2$~dB gains over MAMP with P2P-regular and well-designed P2P-irregular LDPC codes.

\begin{figure*}[!tbp]
	\centering
	\subfigure[$N=500$]{\includegraphics[width=0.465\linewidth]{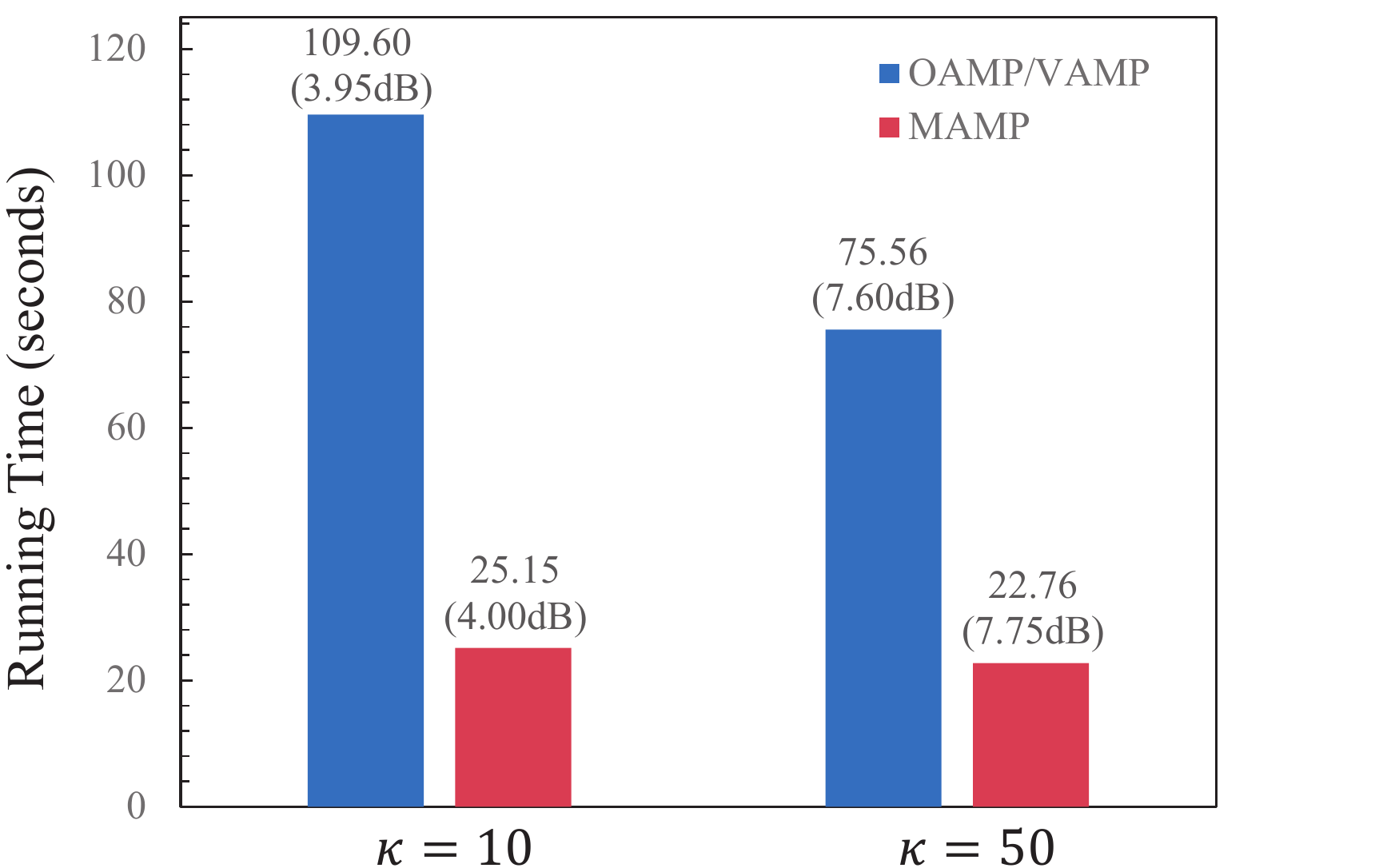}\label{Fig:running time500}}
	\subfigure[$N=5000$]{\includegraphics[width=0.465\linewidth]{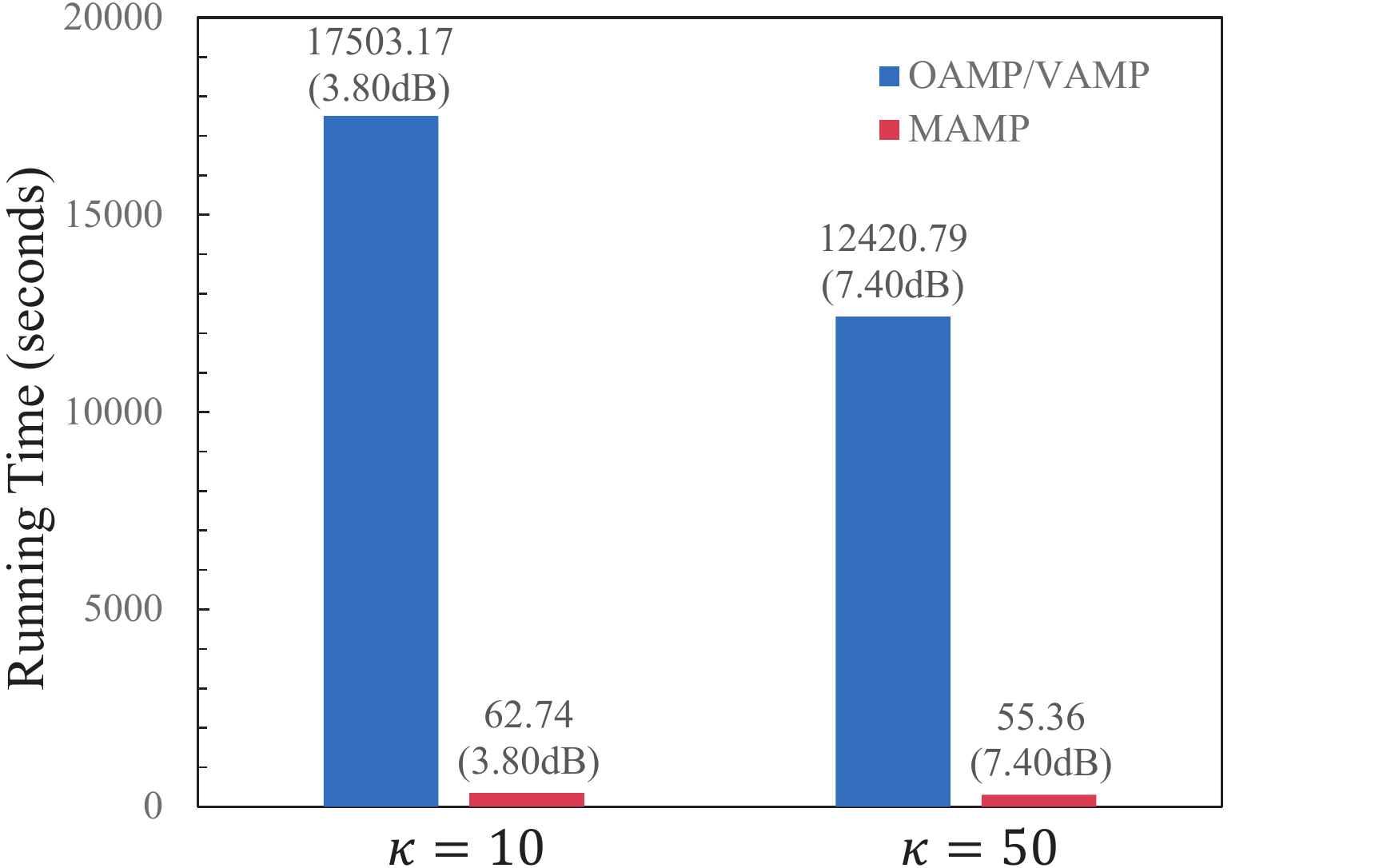}\label{Fig:running time5000}}
	
	\caption{Running time comparison between MAMP and OAMP/VAMP, where target ${\text{BER}}=2\times 10^{-4}$, $\kappa=\{10,50\}$, $\beta=1.5$, and $N=\{500,5000\}$ .}\label{Fig:running time} \vspace{0.5cm}
\end{figure*}

\begin{figure}[!tbp]
	\centering 
	\includegraphics[width=0.93\columnwidth]{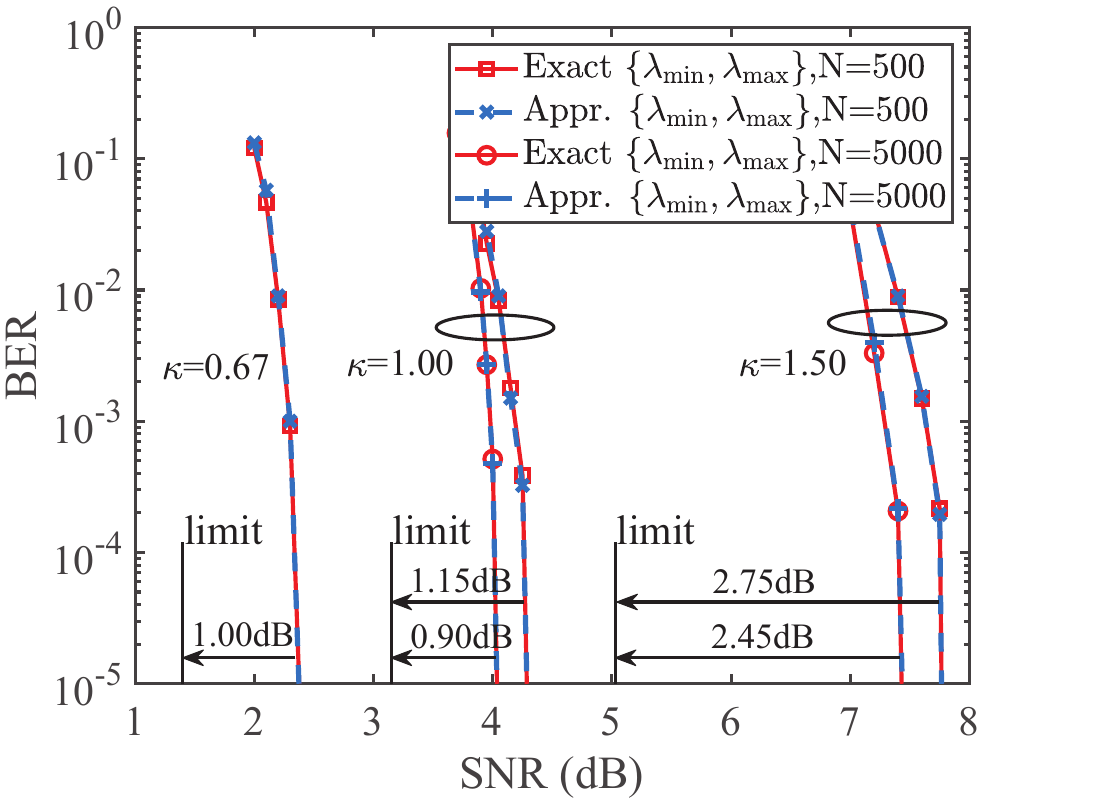} 
	\caption{BER performances of MAMP with exact and approximate (Appr.) $\{\lambda_{\text{min}},\lambda_{\text{max}}\}$ for $\bf{A}_{\rm ill}$, where where $\kappa=50$, $\tau=\{180,140,120\}$ correspond to $\beta=\{0.67,1.00,1.50\}$ for $N=500$, and $\tau=\{160,140\}$ correspond to $\beta=\{1.00,1.50\}$ for $N=5000$, respectively. The optimized LDPC codes are utilized and “limit” is the associated constrained capacity.} \label{Fig:Bound}
\end{figure}

\subsection{Time Complexity Comparison Between MAMP and~OAMP/VAMP}
The time complexity of MAMP and OAMP/VAMP is determined by MLD with complexity $\mathcal{O}(MN\mathcal{T}+N\mathcal{T}^2+\mathcal{T}^3)$ and LD with complexity $\mathcal{O}\left((M^2N+M^3)\mathcal{T}\right)$, respectively~\cite{MAMPTIT}, where $\mathcal{T}$ is the number of iterations.
The time complexity of NLD is identical for MAMP and OAMP/VAMP due to the same demodulation and LDPC decoder employed in NLD. Therefore, compared with OAMP/VAMP, MAMP can achieve the information-theoretic limit of GMIMO with significantly lower time complexity. To intuitively highlight the low-complexity advantage of MAMP, Fig.~\ref{Fig:running time} shows the running time comparison of MAMP and OAMP/VAMP. The running time is obtained by Matlab 2021a on a PC with an 11th Gen Intel Core i7-11700F CPU and 16 GB of RAM.  For $N=500$, Fig.~\ref{Fig:running time500} shows that the running time of MAMP is just $20\%\sim30\%$ of that of OAMP/VAMP. As illustrated in Fig.~\ref{Fig:running time5000}, when $N$ increases to 5000, the running time of MAMP increases by only $145\%$, i.e., $22.76\mr{s} \sim 25.15\mr{s}$ increases to $55.36\mr{s}\sim 62.74 \mr{s}$. In contrast, the running time of OAMP/VAMP increases dramatically by $16,000\%$, i.e., $75.56\mr{s} \sim 109.60\mr{s}$ increases to $12420.79\mr{s}\sim 17503.17\mr{s}$. That is, when $N=5000$, MAMP can achieve the same performances as OAMP/VAMP with $4\text{\textperthousand}$ of the time consumption. Thus, MAMP is the very promising candidate for large-scale systems.

\subsection{Effect of Approximate \texorpdfstring{$\{\lambda_{\text{min}},\lambda_{\text{max}}\}$} {} on MAMP}\label{sec:Bound}
The $\lambda_{\text{min}}$ and $\lambda_{\text{max}}$ of $\bf{A}\bf{A}^{\rm{H}}$ are assumed to be available in the above simulations. If $\lambda_{\text{min}}$ and $\lambda_{\text{max}}$ are unavailable, we can adopt a low-complexity approximation of $\lambda_{\text{min}}$ and $\lambda_{\text{max}}$ provided in~\cite{MAMPTIT}. The detailed approximation steps are shown as follows.

Since $\lambda_{\text{max}}$ is the maximal eigenvalue of $\bf{A}\bf{A}^{\rm{H}}$, $\lambda^{\tau}_{\text{max}}$ is the maximal eigenvalue of $(\bf{A}\bf{A}^{\rm{H}})^{\tau}$, such that $\lambda^{\tau}_{\text{max}} \le \lambda_{\tau} =\mr{tr} \{  (\bf{A}\bf{A}^{\rm{H}})^{\tau} \} $. Due to the right-unitarily-invariant property of $\bf{A}$, $\lambda_{\text{min}}\ge 0$ and $\lambda^{\tau}_{\text{max}} \le \lambda_{\tau}$, hence $\lambda_{\text{min}}$ and $\lambda_{\text{max}}$ can be replaced respectively by a lower bound $\lambda^{\text{low}}_{\text{min}}$ and an upper bound $\lambda^{\text{up}}_{\text{max}}$ as follows:
\BE
    \lambda^{\text{low}}_{\text{min}} = 0 , \;\;\;\;\;\;
    \lambda^{\text{up}}_{\text{max}} = (\lambda_{\tau})^{1/\tau} .
\EE
The $\lambda^{\text{up}}_{\text{max}}$ is tighter for larger $\tau$ and $\lambda_{\tau}$ can be approximated by $\lambda_{\tau} \overset{\text{a.s.}}{=}  \lim\limits_{N \to \infty }  \left \|  \bf{s}_{\tau} \right \|^2$, where 
\BE
    \bf{s}_{\tau} = 
    \left\{ \begin{aligned}
	  &  \bf{A}  (\bf{A}^{\rm{H}}\bf{A})^{\frac{\tau-1}{2}}\bf{s}_0 ,  \;\;\;\;\; \text{if}\; \tau \; \text{is odd}\\
	  &  (\bf{A}^{\rm{H}}\bf{A})^{\frac{\tau}{2}}\bf{s}_0, \;\;\;\;\;\;\;\;\;\; \text{if}\; \tau \; \text{is even}
    \end{aligned} 
    \right. ,
\EE
with $\bf{s}_0 \sim\mathcal{CN}(\mathbf{0}_{N\times 1},\bm{I}_{N\times N})$.
Therefore, the complexity of the above approximation is $\mathcal{O}(MN\tau)$ with $\tau \ll M$ and $N$, especially in large-scale systems. This indicates that the approximation method has no effect on the complexity order of the MLD of MAMP.

As shown in Fig.~\ref{Fig:Bound}, MAMP with approximate eigenvalues can achieve the same~performances as MAMP with exact eigenvalues under different system sizes and channel loads for $\bf{A}_{\rm ill}$, where $\kappa=50$, $\tau=\{180,140,120\}$ correspond to $\beta=\{0.67,1.00,1.50\}$ for $N=500$, and $\tau=\{160,140\}$ correspond to $\beta=\{1.00,1.50\}$ for $N=5000$, respectively. This strongly confirms the effectiveness of the low-complexity approximation for MAMP in coded GMIMO.

\section{Conclusion}
This paper studies the achievable rate analysis and optimal coding principle of the low-complexity MAMP in coded GMIMO, demonstrating the information-theoretic optimality of MAMP. To overcome the difficulty in multi-dimensional SE analysis of MAMP, the fixed-point consistency of MAMP and OAMP/VAMP is proposed to derive the simplified SISO VSE for MAMP, based on which its achievable rate is calculated and optimal coding principle is established to maximize the achievable rate. 
Subsequently, the information-theoretic optimality of MAMP is proved. More importantly, a more general theoretical-analysis equivalence theorem is derived for two arbitrary iterative detection algorithms with the same SE fixed point.
Moreover, the practical optimized LDPC codes are provided for MAMP, where the theoretical decoding thresholds are about $0.3$ dB away from the associated constrained capacities. 
Numerical results show that the finite-length performances of MAMP with optimized LDPC codes are significantly superior to those of MAMP with well-designed P2P LDPC codes. In addition, MAMP can achieve the same performances with $4\text{\textperthousand}$ of running time compared to OAMP/VAMP for large-scale systems.

\appendices

\section{Proof of Lemma~\ref{lem:VSE_MAMP}}\label{APP:VSE_MAMP}

In this proof, the SE and VSE of OAMP/VAMP is firstly reviewed. Then, we prove that MAMP and OAMP/VAMP have the same VSE fixed point. Finally, the VSE of MAMP is derived to calculate the achievable rate of MAMP.

Since the orthogonal transfer functions in the SE of OAMP/VAMP are no longer locally MMSE, the I-MMSE lemma cannot be employed directly for OAMP/VAMP to calculate the achievable rate. Therefore, a VSE of OAMP/VAMP is obtained in \cite{LeiOptOAMP} by incorporating all orthogonal operations into the LD transfer function, such that the NLD transfer function will be local MMSE. Due to the fixed-point equation equivalence of SE and VSE, the VSE is used to analyze the achievable rate of OAMP/VAMP based on I-MMSE lemma. Thus, the VSE of OAMP/VAMP is given as follows.

\emph{VSE of OAMP/VAMP\cite{LeiOptOAMP}:} Let $v_t^{\mr{O}}$ and $\rho_t^{\mr{O}}$ be the input variance and signal to interference plus noise ratio (SINR) of the LD and NLD transfer functions in the VSE of OAMP/VAMP, respectively. Given $snr$ and $\beta$, the VSE of OAMP/VAMP is obtained by \cite[Equnation (51)]{LeiOptOAMP}
\BS\label{Eqn:OAMP-VSE}
\begin{align}
\mathrm{LD:}\;\;\rho_t^{\mr{O}} &= \eta^{\rm{OAMP}}_{\rm{SE}}(v_t^{\mr{O}}) = (v_t^{\mr{O}})^{-1}-[\hat{\gamma}_{\mr{SE}}^{-1}(v_t^{\mr{O}})]^{-1},\label{Eqn:OAMP-VSE-LD}\\
\mathrm{NLD:}\;\;v_{t+1}^{\mr{O}} &= \hat{\phi}^{\mathcal{C}}_{\rm{SE}}(\rho_t^{\mr{O}}),\label{Eqn:OAMP-VSE-NLD}	\end{align}
\ES
where $0\leq \rho_t^{\mr{O}} \leq \rho_{\mr{max}}$ with $\rho_{\mr{max}}$ defined in \eqref{Eqn:CodeMAMP1}, $\hat{\gamma}_{\mr{SE}}(v)= \tfrac{1}{N}{\rm{tr}}\{[snr\bf{A}^{\rm{H}}\bf{A}+v^{-1}\bf{I}]^{-1}\}$ denotes the MSE function of LMMSE detector, and $\hat{\gamma}_{\mr{SE}}^{-1}(\cdot)$ denotes the inverse of $\hat{\gamma}_{\mr{SE}}(\cdot)$.

Assume the VSE fixed point of OAMP/VAMP is $(\rho_*^{\mr{O}}, v_*^{\mr{O}})$. 
Based on the fixed-point equation equivalence of SE and VSE of OAMP/VAMP~\cite{LeiOptOAMP}, we can derive that $\rho_*^{\mr{O}}=1/v_{*}^{\gamma,\rm{O}}$ and $v_*^{\mr{O}}=[(v_{*}^{\gamma,\rm{O}})^{-1}+(v_{*}^{\phi,\rm{O}})^{-1}]^{-1}$, where $(v_{*}^{\gamma,\rm{O}}, v_{*}^{\phi,\rm{O}})$ is the SE fixed point of OAMP/VAMP. For MAMP in Fig.~\ref{Fig:MAMP receiver1}, assume the VSE fixed point is $(\rho_*^{\gamma}, v_*^{\hat{\phi}})$, where $\rho_*^{\gamma}=1/v^{\gamma}_{*}$. Due to the orthogonalization, we have 
$v^{\phi}_{*} =  [(v^{\hat{\phi}}_{*})^{-1} - (v^{\gamma}_{*})^{-1}]^{-1}$, i.e., $v^{\hat{\phi}}_{*} =  [(v^{\gamma}_{*})^{-1} + (v^{\phi}_{*})^{-1}]^{-1}$.
Meanwhile, based on Lemma~\ref{lem:same_fp}, for fixed $\hat{\phi}^{\mathcal{C}}_{\rm{SE}}(\cdot)$, MAMP and OAMP/VAMP converge to the same SE fixed point, i.e., $(v_{*}^{\gamma}, v_{*}^{\phi})=(v_{*}^{\gamma,\rm{O}}, v_{*}^{\phi,\rm{O}})$.
Therefore, MAMP and OAMP/VAMP have the same VSE fixed point, i.e., 
\BE\label{Eqn:vse_sf}
(\rho_*^{\gamma}, v_*^{\hat{\phi}})=(\rho_*^{\rm{O}}, v_*^{\rm{O}}). 
\EE

\begin{figure}[!tbp]
	\centering
	\includegraphics[width=0.8\linewidth]{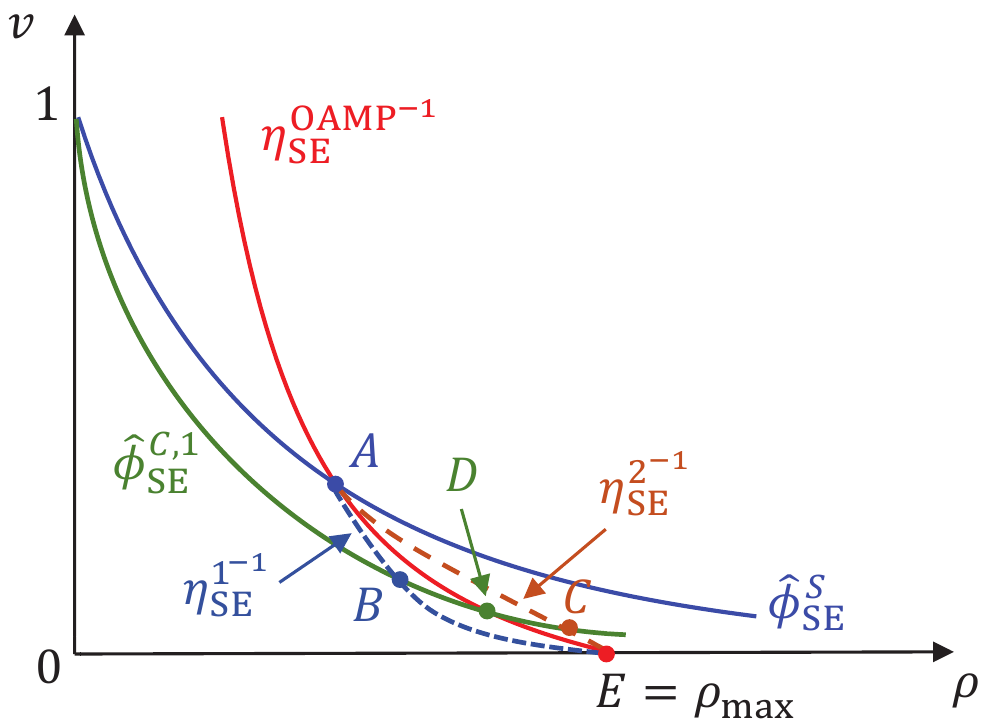}
	
	\caption{Graphical illustration of VSE for OAMP/VAMP and MAMP, where $\eta_{\mr{SE}}^{\mr{OAMP}^{-1}}(\cdot)$ is the inverse of $\eta_{\mr{SE}}^{\mr{OAMP}}(\cdot)$ and $\hat{\phi}^{\mathcal{S}}_{\rm{SE}}(\cdot)$ is the MMSE function of demodulation. MAMP and OAMP/VAMP have the same VSE fixed point~A. $\eta^{1^{-1}}_{\rm{SE}}(\cdot)$ and $\eta^{2^{-1}}_{\rm{SE}}(\cdot)$ are two candidate inverse functions of VSE transfer functions of MAMP's MLD. Given an MMSE function $\hat{\phi}^{\mathcal{C},1}_{\rm{SE}}(\cdot)$ of decoder, its intersections with $\eta^{1^{-1}}_{\rm{SE}}(\cdot)$, $\eta^{2^{-1}}_{\rm{SE}}(\cdot)$, and $\eta_{\mr{SE}}^{\mr{OAMP}^{-1}}(\cdot)$ are point B, C, and D, respectively.}\label{Fig:VSE-curve}
\end{figure}

As shown in Fig.~\ref{Fig:VSE-curve}, assume that there is a unique VSE fixed point A $=(\rho_*^{\mr{O}}, v_*^{\mr{O}})$ between $\eta^{\rm{OAMP}^{-1}}_{\rm{SE}}(\cdot)$ and $\hat{\phi}^{\mathcal{S}}_{\rm{SE}}(\cdot)$ in uncoded GMIMO, where $\hat{\phi}^{\mathcal{S}}_{\rm{SE}}(\cdot)$ is the MMSE function of demodulation. 
Based on \eqref{Eqn:vse_sf}, the VSE fixed point between $\eta^{-1}_{\rm{SE}}(\cdot)$ and $\hat{\phi}^{\mathcal{S}}_{\rm{SE}}(\cdot)$ for MAMP is also point A~$=(\rho_*^{\gamma}, v_*^{\hat{\phi}})$. When $0\le\rho^{\gamma}<\rho^{\gamma}_{*}$, MAMP can iterate with $\hat{\phi}^{\mathcal{S}}_{\rm{SE}}(\rho^{\gamma})<\eta_{\rm{SE}}^{-1}(\rho^{\gamma})$. Furthermore, $\hat{\phi}^{\mathcal{C}}_{\rm{SE}}(\rho^{\gamma})<\hat{\phi}^{\mathcal{S}}_{\rm{SE}}(\rho^{\gamma})$ is obtained because of the coding gain~\cite{LeiOptOAMP}.
Therefore, $\hat{\phi}^{\mathcal{C}}_{\rm{SE}}(\rho^{\gamma})<\hat{\phi}^{\mathcal{S}}_{\rm{SE}}(\rho^{\gamma})<\eta_{\rm{SE}}^{-1}(\rho^{\gamma})$, i.e., $\hat{\phi}^{\mathcal{C}}_{\rm{SE}}(\rho^{\gamma})$ is limited to $\hat{\phi}^{\mathcal{S}}_{\rm{SE}}(\rho^{\gamma})$, and it is reasonable to ignore the impact of the specific expression of $\eta^{-1}_{\rm{SE}}(\rho^{\gamma})$ on $\hat{\phi}^{\mathcal{C}}_{\rm{SE}}(\rho^{\gamma})$.
Therefore, we only focus on the expression of $\eta_{\rm{SE}}^{-1}(\rho^{\gamma})$ for $\rho^{\gamma}\ge\rho^{\gamma}_{*}$.

Given $\forall \rho_1,\rho_2 \in[\rho^{\gamma}_{*},\rho_{\rm max}]$ with $\rho_1<\rho_2$, assume $\eta_{\rm{SE}}^{-1}(\rho^{\gamma})>\eta^{\rm{OAMP}^{-1}}_{\rm{SE}}(\rho^{\gamma})$ for $\rho^{\gamma}\in [ \rho_1,\rho_2]$. Given an MMSE function $\hat{\phi}^{\mathcal{C},1}_{\rm{SE}}(\cdot)$ of decoder, due to the coding gain~\cite{LeiOptOAMP}, $\hat{\phi}^{\mathcal{C},1}_{\rm{SE}}(\cdot)<\hat{\phi}^{\mathcal{S}}_{\rm{SE}}(\cdot)$ is obtained. Since MMSE function is monotone decreasing, $\hat{\phi}^{\mathcal{C},1}_{\rm{SE}}(\rho^{\gamma})$ has two different VSE fixed point with $\eta_{\rm{SE}}^{-1}(\rho^{\gamma})$ and $\eta^{\rm{OAMP}^{-1}}_{\rm{SE}}(\rho^{\gamma})$, which contradicts \eqref{Eqn:vse_sf}.
Similarly, assume $\eta_{\rm{SE}}^{-1}(\rho^{\gamma})<\eta^{\rm{OAMP}^{-1}}_{\rm{SE}}(\rho^{\gamma})$ for $\rho^{\gamma}\in [ \rho_1,\rho_2]$. There are still two different VSE fixed point with $\eta_{\rm{SE}}^{-1}(\rho^{\gamma})$ and $\eta^{\rm{OAMP}^{-1}}_{\rm{SE}}(\rho^{\gamma})$, which contradicts \eqref{Eqn:vse_sf}. As a result, due to the arbitrariness of $\rho_1$ and $\rho_2$, $\eta_{\rm{SE}}^{-1}(\rho^{\gamma})=\eta^{\rm{OAMP}^{-1}}_{\rm{SE}}(\rho^{\gamma})$ (i.e., $\eta_{\rm{SE}}(\rho^{\gamma})=\eta^{\rm{OAMP}}_{\rm{SE}}(\rho^{\gamma})$) for $\rho^{\gamma}\in[\rho^{\gamma}_{*},\rho_{\rm max}]$ is proved. Since the specific expression of $\eta_{\rm{SE}}(\rho^{\gamma})$ for $\rho^{\gamma}\in[0,\rho^{\gamma}_{*})$ is negligible, we let $\eta_{\rm{SE}}(\rho^{\gamma})=\eta^{\rm{OAMP}}_{\rm{SE}}(\rho^{\gamma})$ with $\rho^{\gamma}\in[0,\rho_{\rm max}]$ for simplicity.
Therefore, the VSE of MAMP is derived, which is presented in Lemma~\ref{lem:VSE_MAMP}.

To illustrate this proof intuitively, letting $\rho_1=\rho^{\gamma}_{*}$ and $\rho_2=\rho_{\rm max}$, two candidate functions $\eta^{1^{-1}}_{\rm{SE}}(\cdot)$ and $\eta^{2^{-1}}_{\rm{SE}}(\cdot)$ of $\eta^{-1}_{\rm{SE}}(\cdot)$ are obtained in Fig.~\ref{Fig:VSE-curve}. The intersections of $\hat{\phi}^{\mathcal{C},1}_{\rm{SE}}(\cdot)$ with $\eta^{1^{-1}}_{\rm{SE}}(\cdot)$, $\eta^{2^{-1}}_{\rm{SE}}(\cdot)$, and $\eta^{\rm{OAMP}^{-1}}_{\rm{SE}}(\cdot)$ are point B, C, and D, respectively, which contradicts \eqref{Eqn:vse_sf}. Therefore, we can derive that $\eta_{\rm{SE}}(\cdot)=\eta^{\rm{OAMP}}_{\rm{SE}}(\cdot)$.

\footnotesize
\bibliographystyle{IEEEtran}
\bibliography{manuscript}

\end{document}